\documentclass[fleqn,usenatbib]{mnras}
\usepackage{hyperref}
\hypersetup{
    colorlinks=true,
    linkcolor=blue,
    filecolor=magenta,
    urlcolor=cyan,
    pdfpagemode=FullScreen,
    }

\usepackage{newtxtext,newtxmath}
\usepackage[T1]{fontenc}
\usepackage{ae,aecompl}
\usepackage{graphicx}
\usepackage{amsmath}
\usepackage{lipsum}

\usepackage{multirow}
\usepackage{dcolumn}

\def\lsim{~\rlap{$<$}{\lower 1.0ex\hbox{$\sim$}}}
\def\gsim{~\rlap{$>$}{\lower 1.0ex\hbox{$\sim$}}}

\title[Evolution of the black hole to host mass ratio]{Diagnosing the massive-seed pathway to high-redshift black holes: statistics of the evolving black hole to host galaxy mass ratio}
\author[M.~T.~Scoggins et al]{Matthew~T.~Scoggins,$^{1}$\thanks{E-mail: mts2188@columbia.edu}
\ and Zolt{\'{a}}n~Haiman,$^{1,2}$
\\
$^{1}$Department of Astronomy, Columbia University, New York, NY, 10027\\
$^{1}$Department of Physics, Columbia University, New York, NY, 10027\\
}

\date{Accepted 2024 June 03. Received 2024 May 10; in original form 2023 September 29}

\pubyear{2024}

\begin{document}
\label{firstpage}
\pagerange{\pageref{firstpage}--\pageref{lastpage}}
\maketitle

\begin{abstract}
    Supermassive black holes (SMBHs) with masses of $\sim 10^9 {\rm M_\odot}$ within the first billion year of the universe challenge our conventional understanding of black hole formation and growth. One pathway to these SMBHs proposes that supermassive stars (SMSs) born in pristine atomic cooling haloes (ACHs) yield massive seed BHs evolving to these early SMBHs. This scenario leads to an overly massive BH galaxy (OMBG), in which the BH to stellar mass ratio is initially $M_{\rm bh}/M_* \geq 1$, well in excess of the typical values of $\sim 10^{-3}$ at low redshifts. Previously, we have investigated two massive seed BH candidates from the \texttt{Renaissance} simulation and found that they remain outliers on the $M_{\rm bh}-M_{*}$ relation until the OMBG merges with a much more massive halo at $z{=}8$. In this work, we use Monte-Carlo merger trees to investigate the evolution of the $M_{\rm bh}-M_{*}$ relation for $50,000$ protogalaxies hosting massive BH seeds, across $10,000$ trees that merge into a $10^{12} {\rm M_\odot}$ halo at $z{=}6$. We find that up to $60\%$ (depending on growth parameters) of these OMBGs remain strong outliers for several 100 Myr, down to redshifts detectable with {\it JWST} and with sensitive X-ray telescopes. This represents a way to diagnose the massive-seed formation pathway for early SMBHs.  We expect to find ${\sim} 0.1{-}1$ of these objects per {\it JWST} NIRCam field per unit redshift at $z\gsim 6$. Recently detected SMBHs with masses of $\sim 10^7~{\rm M_\odot}$ and low inferred stellar-mass hosts may be examples of this population. 
\end{abstract}

\section{Introduction}
\label{sec:intro}

There are over 200 detections of bright quasars powered by supermassive black holes (SMBHs) with masses on the order of $10^9~{\rm M_\odot}$ at redshift $z \geq 6$ (for recent compilations, see \citealt{Inayoshi_2020,Bosman_2022,Fan_2023}). The existence of these SMBHs with ages $\leq1$ Gyr challenges our conventional understanding of black hole formation and growth. While Eddington-limited accretion throughout the entire assembly history of these black holes is unlikely, some observations suggest masses that require even higher average accretion rates sustained throughout the (then) age of the universe.

Several formation pathways have emerged that attempt to explain these SMBHs. Most of these pathways fall into two categories, with so-called light and heavy seeds. Light-seed models propose a Population III (hereafter Pop~III) stellar remnant black hole that grows at at least modestly super-Eddington rates for a significant fraction of its life (e.g. \citealt{Tanaka_2009, Volonteri_2010}). This is necessary for a $10-100{\rm M_\odot}$ seed to reach $10^9 {\rm M_\odot}$ in less than 1 Gyr. Heavy-seed models invoke one of several mechanisms that rapidly produce a $10^4-10^6 {\rm M_\odot}$ seed black hole, which then grows at the Eddington limit. Mechanisms producing heavy seeds include hyper-Eddington accretion onto a lower-mass BH \citep{Ryu_2016, Inayoshi_2016}, runaway collisions between stellar-mass BHs and/or stars in dense proto-clusters \citep{Boekholt_2018, Tagawa_2020, Escala_2021, Vergara_2022, Schleicher_2022}, and the so-called direct-collapse black hole (DCBH) scenario \citep{Agarwal_2012, Latif_2013, Ferrara_2014, Inayoshi_2014,Sugimura_2014,Tanaka_2014,Becerra_2015, Hosokawa_2016,Chon_2016, Umeda_2016, Hirano_2017, Haemmerle_2018}. Hyper-Eddington accretion would allow a small BH to quickly become a $10^{5-6} {\rm M_\odot}$ seed, while runaway mergers in a primordial star cluster could quickly give rise to a $10^{4-5} {\rm M_\odot}$ seed. The most studied heavy-seed scenario, direct-collapse, proposes that chemically pristine haloes that reach the atomic cooling threshold (ACT), without prior star formation, collapse via rapid atomic (hydrogen) cooling and form a supermassive star (SMS). Reaching the atomic-cooling halo (ACH) stage without prior fragmentation, star-formation, and metal-enrichment can be achieved via several mechanisms that prevent or offset cooling. Intense Lyman-Werner (LW) radiation can dissociate ${\rm H_2}$ and prevent ${\rm H_2}$ cooling, haloes can experience dynamical heating through rapid halo mergers, and large residual baryonic streaming motions from recombination can prevent gas infall and contraction in low-mass DM "minihaloes".

All of the mechanisms that lead to heavy seeds share an interesting feature, resulting from the lack of prior star formation or little remaining stellar mass at the time of black hole formation: the mass of the black hole seed is initially comparable to or much greater than the surrounding stellar mass, $M_{\rm bh}/M_* \geq 1$.  These so-called overly massive black hole galaxies (OMBGs) are unusual compared to massive black holes at low redshifts, which reside in much more massive stellar hosts with $M_{\rm bh}/M_* \sim 10^{-3}$, or even compared to recent observations of SMBHs and their host galaxies at $z\approx 6$, which appear to have a somewhat higher ratio, $M_{\rm bh}/M_* \sim 10^{-2}$ \citep{Pacucci_2023}.  {\it JWST} has recently enabled the detection of several high-redshift lower-mass SMBHs.   Establishing their place on the $M_{\rm bh}/M_*$ relation would help determine the origin of these SMBHs. See \S~\ref{sec:discussion} for a brief compilation of some of these recently detected black holes and a discussion of where they stand in the BH-host galaxy mass relation.

In \citet[][hereafter S22]{Scoggins_2022}, we investigated the DCBH pathway, where a black hole seed of $10^4-10^6 {\rm M_\odot}$ forms in the early universe and grows via Eddington-limited accretion into the $>10^9 {\rm M_\odot}$ SMBHs we observe today. We focused on two candidate DCBHs identified in a suite of cosmological radiation-hydrodynamic and N-body simulations, the \texttt{Renaissance} simulations \citep{OShea_2015, Xu_2016}.  These DCBH candidates were found in the most massive halo (MMH) and the halo which saw the highest Lyman-Werner flux (LWH). Although their $M_{\rm bh}/M_*$ ratio is initially extremely high, internal star-formation and mergers with other haloes with typical $M_{\rm bh}-M_*$ relations subsequently drive this ratio to approach $\geq 10^{-2}$. Our goal in S22 was to follow the merger histories of these two DCBH host candidate haloes in the underlying \texttt{Renaissance} N-body simulations, and to assess how long their $M_{\rm bh}/M_*$ ratio might remain outstandingly high. We found that with either Eddington-limited growth or a super-Eddington prescription \citep{Hu_2022a, Hu_2022b}, both candidates satisfy $M_{\rm bh}/M_* \gtrsim 1$ until they experience a merger with a much more massive ($\sim10^{11}~{\rm M_\odot}$) halo, which happened near $z{\sim} 8$ in both cases. 

A key insight gained in S22 was that the mass relation is not efficiently normalized by minor mergers, but only by mergers with much more massive haloes. In the present work, we follow up on this earlier study, and generate $10^4$ Monte-Carlo halo merger trees, each representing the history of a $M_{\rm halo} = 10^{12} {\rm M_\odot}$ dark matter (DM) halo at redshift $z{=}6$.  We then search for DCBH candidate sites within these trees, and track their mass-relation evolution in a way similar to S22.    Our goal is to characterize the statistics of how long the DCBHs remain outliers in the BH-host mass relations.  This allows us to determine how typical or atypical the MMH and LWH were, and whether the over-massive relation lifetime (hereafter OMRL) -- the duration for which a newly-born DCBH and its stellar host have a mass ratio $M_{\rm bh}/M_*$ above some pre-specified minimum value --  is long enough to be uncovered by observations at $z\gsim 8$ where these early SMBHs are detected.

The rest of this paper is organised as follows. In \S~\ref{sec:methods} we describe our Monte-Carlo merger trees, our selection of DCBH sites, and our simple models for the evolving black hole and stellar masses. In \S~\ref{sec:results} we present our results on the DCBH candidates and the distribution of their OMRLs.
In \S\ref{sec:discussion} we discuss the possibility of detecting OMBGs and using them to diagnose the massive-seed pathway. Finally, we summarise our findings and offer our conclusions in~\S~\ref{sec:conclusion}.

\begin{figure*}   
\includegraphics[width=1.8\columnwidth]{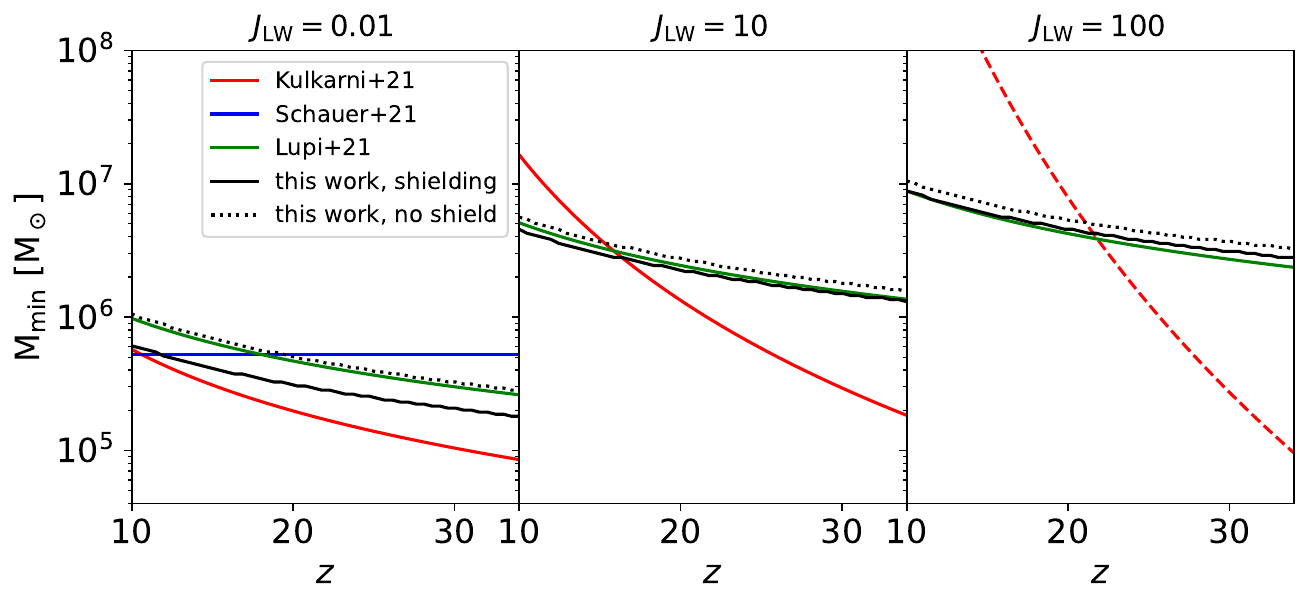}
\caption{A comparison of several models for the minimum mass required for cooling and collapse of gas in primordial haloes. Two models derive this minimum mass by identifying haloes undergoing collapse in cosmological simulations with varying $J_{\rm LW}$ backgrounds, \citet{Kulkarni_2021} (red, $J_{\rm LW}\ \epsilon \{0,1,10, 30\}$)  and \citet{Schauer_2021} (blue, $J_{\rm LW}\ \epsilon \{0,0.1,0.01\}$). \citet{Lupi_2021} (green) uses an analytical model similar to ours, but we also include a model that accounts for self-shielding. Our full model will estimate evolution-dependent minimum masses, where we also include dynamical heating.
}
    \label{fig:mcrit}
\end{figure*}

\section{Methods}
\label{sec:methods}

In this section we summarise the methods used to generate our Monte-Carlo merger trees, the criteria to select massive DCBH seed candidates, and the prescriptions for black hole growth and mergers. All of the analysis used in this work assumes the following cosmological parameters: $\Omega_{\Lambda} = 0.693$, $\Omega_m = 0.307$, $\Omega_b=0.0486$, $\sigma_8=0.81$, and $h =0.67$ \citep{Planck_2018}.

\subsection{Monte-Carlo merger trees}

We generate dark matter halo histories using Monte-Carlo merger trees based on the Extend Press-Schechter theory \citep{Press_1974}, following the algorithm detailed in \citet{Parkinson_2007}, which is a modification of the algorithm used in the \texttt{GALFORM} semi-analytic galaxy formation model \citep{Cole_2000}. We generate $10^4$ merger trees with a parent mass of $10^{12} {\rm M_\odot}$ at redshift $z=6$, and a redshift step size of $dz=0.15$. We impose a mass resolution of $10^{5} {\rm M_\odot}$ which also determines the highest redshift at which branches of the merger trees terminate, typically at $z_{\rm max}\approx 30-35$.

\subsection{Identifying massive BH seed sites}

A 'direct-collapse' black hole can be achieved via an intermediary ${\sim} 10^5 {\rm M_\odot}$ SMS. In order to form such a supermassive star, gas must reach atomic cooling ($T_{\rm vir} \sim 10^4{\rm K}$), where runaway atomic cooling processes allow isothermal collapse, avoiding fragmentation and instead forming a large central SMS. Alternative models to produce massive BH seeds similarly require pristine gas in ACHs (see \S~\ref{sec:intro}). The gas in most haloes begins to cool and collapse before reaching the ACT. ${\rm H_2}$ plays the primary role in this collapse, where a large ${\rm H_2}$ abundance can rapidly radiate energy out of the halo, leading to cooling and fragmentation. There are several processes that influence the cooling rate: (i) Lyman-Werner radiation (with specific intensity $J_{\rm LW}$) from a neighboring galaxy, or, in the case of mini-haloes, background LW radiation \citep{Dijkstra_2008, Dijkstra_2014} can dissociate ${\rm H_2}$ and slow or completely stop cooling \citep{Haiman_1997}, (ii) dynamical heating (at a rate $\Gamma_{\rm dyn}$) from rapid halo mergers can efficiently heat the halo and offset cooling \citep{Yoshida_2003, Wise_2019}, and (iii) large baryonic streaming motions ($v_{\rm stream}$) can prevent gas infall and contraction in DM haloes \citep{Greif_2011,Latif_2014}. (iv) Local infrared (IR) sources can also stunt ${\rm H_2}$ formation by photo-detaching ${\rm H}^-$, which is an intermediary needed to form ${\rm H_2}$ \citep{Wolcott_2012}. Finally, (v) X-rays can ionize neutral hydrogen, creating free electrons which increase the ${\rm H}^-$ abundance, in turn increasing ${\rm H_2}$ abundance \citep{Haiman_1996}, while X-rays can also warm the intergalactic medium and suppress the formation and growth of subsequent generations of BHs \citep{Tanaka_2012}. If these processes can prevent or offset ${\rm H_2}$ cooling as 
the halo grows to the atomic cooling stage with $T_{\rm vir}\sim 10^4{\rm K}$, the emission of atomic hydrogen will rapidly cool the halo, allowing for isothermal collapse, possibly producing a massive BH seed via a SMS or through one of the alternative scenarios described in \S~\ref{sec:intro}.

To apply these criteria at each halo in every merger tree, we compare the cooling time $t_{\rm cool}$ to the Hubble time $t_{\rm hub}$, where a halo becomes the host of a massive BH seed if none of the progenitors of that halo had experienced prior star formation, i.e. $t_{\rm cool} > t_{\rm hub}$ throughout the history of each progenitor. Our calculation for the Hubble time follows
\begin{align*}
    t_{\rm hub} = \frac{2}{3\sqrt{\Omega_\Lambda}} \ln(b + \sqrt{1+b^2})
\end{align*}
where $b = \sqrt{\Omega_\Lambda/\Omega_m}(z+1)^{-1.5}$. The cooling time follows $t_{\rm cool} = u/(\Lambda_{\rm cool} n_{\rm H}n_{\rm H_2} - \Gamma_{\rm dyn})$ for energy density $u =\frac{3}{2} n_{\rm gas} k T $, cooling rate $\Lambda_{\rm cool}$, and heating rate $\Gamma_{\rm dyn}$. The cooling rate is given by equation (A.2) of \citet{Galli_1998},
\begin{align}
    \Lambda = \frac{\Lambda({\rm LTE})}{ 1 + [n^{\rm cr}/n({\rm H})]},
\end{align}
where $\Lambda({\rm LTE})$ is the LTE cooling function of \citet{Hollenbach_1979}, and $n^{\rm cr}/n({\rm H})$ follows $\frac{\Lambda({\rm LTE})}{\Lambda({\rm n_H} \rightarrow 0)}$ for the low-density limit of the cooling function. This is well approximated by equation (A.7) of \citet{Galli_1998}.  For dynamical heating, we follow equation (1) of \citet{Wise_2019}, which is similar to equation (3) of \citet{Yoshida_2003},
\begin{align}
    \Gamma_{\rm dyn} = \frac{T_{\rm halo}}{M_{\rm halo}} \frac{k_{\rm B}}{\gamma -1} \frac{dM_{\rm halo}}{dt},
\end{align} 
for adiabatic index $\gamma = 5/3$. We assume in the absence of cooling the gas compresses adiabatically, giving a maximum central number density $n_c {\sim} 6 (\frac{T_{\rm vir}}{1000 {\rm K}})^{\frac{3}{2}}$ cm$^{-3}$ \citep{Visbal_2014}, $T_{\rm vir}$ from equation (26) of \citet{Barkana_2001}, and total number density $n {=} f_{\rm gas} n_c$ with scaling factor $f_{\rm gas}$. See below for a discussion of $f_{\rm gas}$. We approximate the ${\rm H_2}$ abundance assuming ${\rm H_2}$ dissociation via LW radiation is in equilibrium with ${\rm H_2}$ formation via  ${\rm H + e^- \rightarrow {\rm H}^- + hv}$ followed by ${\rm H + {\rm H}^- \rightarrow H_2 + e^-}$,
\begin{align}
    n_{\rm H_2} = k_9 n_{\rm H} n_{\rm e}/k_{\rm LW}
\end{align}
with $k_9$ given in Table (A1) of \citet{Oh_2002} and the post-recombination residual electron fraction $n_{\rm e}/n_{\rm H} = 1.2\times 10^{-5} \sqrt{\Omega_m}/(\Omega_b h)$ \citep{Peebles_1993}. The dissociation rate by Lyman-Werner radiation is approximated by
$k_{\rm LW} = 1.39 \times 10^{-12} J_{\rm LW}~{\rm s^{-1}}$ for LW specific intensity $J_{\rm LW}$ in units $10^{-21}$ erg cm$^{-2}$ s$^{-1}$ Hz$^{-1}$ sr$^{-1}$ \citep{Wolcott_2017} .

\subsection{Lyman-Werner Radiation}
Though our merger histories lack any spatial information, we can calculate the mean LW flux seen by a halo following the model implemented in \citet{Dijkstra_2014} and \citet{Li_2021}. The average number of haloes within the mass range $m \pm dm/2$ in a spherical shell of radius $r$ and thickness $dr$ is given by
\begin{align}
    \frac{dN(m,r)}{dmdr}dmdr = 4\pi r^2 dr  (1+z)^3 \frac{dn_{\rm ST}(m,z)}{dm}dm [1 + \xi(M,m,z,r)]
\end{align} 
where $dn_{\rm ST}(m,z)/dm$ is the modified Press-Schechter mass function \citep[see eq.~5 of ][]{Sheth_2001} 
and $\xi(M,m,z,r)$ is the two-point halo correlation function, giving the excess probability of finding a halo of mass $m$ at distance $r$ from a halo of mass $M$ \citep{Iliev_2003}. Using this, we calculate the mean Lyman-Werner radiation imparted on a halo of mass $M_{\rm halo}$ at redshift $z$ as
\begin{align}
    \overline{J}_{\rm LW}(M_{\rm halo}, z) = \int_{m_{\rm min}}^{m_{\rm max}} \int_{r_{\rm min}}^{r_{\rm max} }  \frac{dN(m,r)}{dmdr} \frac{L_{\rm LW}}{16 \pi^2 r^2}  dmdr
    \label{eqn:j_mean}
\end{align} 
for LW luminosity $L_{\rm LW}$.  Note that $L_{\rm LW}=L_{\rm LW}(m,z)$ depends on the redshift and mass of each neighboring halo, with stellar mass $m_*=m_*(m,z)$ assigned to each halo as described below. See \citet{Li_2021} for the details of the integration bounds and LW luminosity per stellar mass.

We find $\overline{J}_{\rm LW} < 100$ for most haloes in the progenitors in our $10^4$ merger trees (though $\overline{J}_{\rm LW}$ can exceed $100$ at $z\gtrsim15$ for some haloes, see fig. 2 of \citealt{Li_2021}) while the sites that form DCBHs have conventionally required much larger LW intensities ($J_{\rm crit} \sim 10^3$; see, e.g. \citealt{Shang_2010}, \citealt{Agarwal_2016}, \citealt{Glover_2015} or \citealt{Wolcott_2017}). This is due to equation~\ref{eqn:j_mean} capturing the mean Lyman-Werner radiation, where the LW intensity distribution, due to stochastic variations in the spatial distribution of nearby haloes, is not included. 

To capture this scatter, we draw from a numerically determined $J_{\rm LW}$ probability distribution shown in Fig. 9 of \citet{Lupi_2021}, with some simplifications. For a halo with mass $M_{\rm halo}$ at redshift $z$, the distribution is approximated as symmetric and centered on $c = \log_{10}(\overline{J_{\rm LW}}(M_{\rm halo}, z))$ (where the median (peak) is approximately equal to the mean for a distribution that is symmetric in log space with evenly spaced bins). Letting $x= \log_{10}(J_{\rm LW})$, the distribution describing the number of haloes, $N_{\rm halo}$, experiencing $x$ follows
\begin{align}
\label{eqn:n_halo_dist}
    \log(N_{\rm halo}(x)) = A - 2|x-c|
\end{align}
for normalization $A$. We assume the distribution is within 5 orders of magnitude from the peak, $|x-c| \leq 5$, though increasing this range and allowing broader tails has negligible effects on the results.
While the $J_{\rm LW}$ distribution of the pristine DH candidates in \citet{Lupi_2021} is not quite symmetric, our $\overline{J}_{\rm LW}$ values are typically $< 10^2$ whereas their peak is at $> 10^2$, meaning our distribution tends to be conservative with $J_{\rm LW}$ predictions. 

For each halo above the ACT (with $T_{\rm vir} \gtrsim 10^4 {\rm K}$), we calculate $\overline{J_{\rm LW}}(M_{\rm halo}, z)$ and draw a value $J_{\rm draw}$ from the distribution described in Eq.~\ref{eqn:n_halo_dist}. For a halo just above the ACT, we calculate $\alpha = J_{\rm draw}/\overline{J}_{\rm LW}$, and propagate this ratio down the branches of the tree (towards higher $z$). This means that a minihalo below the ACT which eventually merges into an ACH with a particular value of $\alpha$ had been historically exposed to a LW flux of $J_{\rm LW}(M_{\rm halo},z) = \alpha \overline{J}_{\rm LW}(M_{\rm halo},z)$ at earlier redshifts. Our simple treatment above attempts to account for the fact that a halo experiencing an unusually high (low) LW flux is in an overcrowded (underdense) region, and presumably the progenitors of these haloes likewise will be exposed to higher (lower) LW fluxes compared to the average flux for a halo with that mass at that time. While we assume here that $\alpha$ remains fixed, $\alpha$ for a given halo may evolve with redshift.  Since overdensities generally grow over time, it is possible that the effective $\alpha$ tends grow over time as well, implying that fixing $\alpha$ may lead to an overestimation of $J_{\rm LW}$ at earlier times.  We leave it to future, 3D cosmological simulation, to estimate how $\alpha$ may typically evolve (and how its evolution varies from halo to halo).

Our work accounts for the two primary mechanisms that offset cooling, ${\rm H_2}$ dissociation via Lyman-Werner radiation and heating through mergers. While ${\rm H_2}$ dissociation via Lyman-Werner radiation is thought to play the primary role, there is disagreement in simulations on exactly when they lead to collapse \citep{Schauer_2021, Kulkarni_2021}. To highlight this, we compare our model (excluding the effects of dynamical heating) to three other models, shown in Fig.~\ref{fig:mcrit}. Here, we show two formulae derived from cosmological simulations, where \citet{Schauer_2021} and \citet{Kulkarni_2021} both define criteria for halo collapse and follow primordial haloes through a cosmological simulation. They both fit the point of collapse as a function of redshift and LW flux, with \citet{Kulkarni_2021} fitting for $0\leq J_{\rm LW} \leq 30$  and \citet{Schauer_2021}  fitting for $0\leq J_{\rm LW} \leq 0.1$. Both works also include the effects of baryonic streaming motions, which we have set to zero in our comparison. The desire to account for dynamical heating via mergers, which plays an important role in the creation of these rare DCBH sites, prevents us from applying these models.
Further, the required value of $J_{\rm LW}$ which typically leads to the creation of DCBHs is outside the range of these fitting formulae. Our analytic model, which is very similar to \citet{Lupi_2021}, allows us to account for dynamical heating and does not diverge for large values of $J_{\rm LW}$. 

Comparison of our model with the three models previously discussed motivates us to set $f_{\rm gas} {=} 0.2$. Selecting $f_{\rm gas} {=} 0.2$ sets the predictions for our model (with $J_{\rm LW} = 0.01$) to be bounded by the other models across $6 \leq z \leq 50$. As discussed in \citet{Lupi_2021}, setting $f_{\rm gas} {=} 1$ improves the agreement with \citet{Kulkarni_2021} (and worsens agreement with \citealt{Schauer_2021}), though this only decreases the minimum mass required for collapse of primordial haloes by a factor of ${\sim} 2$.

Finally, not all of the Lyman-Werner radiation reaches the center of the halo, where self-shielding effects reduce the total radiation seen by the core of the halo. To capture this effect, we use the self-shielding fitting formula from \citet{Wolcott_2019}, which calculates the fraction of the incident radiation that passes through a column of ${\rm H_2}$:
\begin{align}
    f_{\rm shield} &= \frac{0.965}{(1+x/b_5)^{\alpha(n,T)}} + \frac{0.035}{(1+x)^{0.5}} \exp[-8.5\times 10^{-4}   (1+x)^{0.5}]\\
    \alpha(n,T) &= A_1(T) \exp(-c_1 \times\log(n/{\rm cm^{-3}})) + A_2(T)\\
    A_1(T) &= c_2 \times \log(T/{\rm K}) - c_3\\
    A_2(T) &= -c_4 \times \log(T/{\rm K}) + c_5
\end{align}
with $c_1 =0.2856$, $c_2 = 0.8711$, $c_3 = 1.928$, $c_4 = 0.9639$, $c_5 = 3.892$, $x = N_{{\rm H}_2}/5\times 10^{14}$ cm$^{-2}$, $b_5 = b/10^5$ cm s$^{-1}$ and $b$ the Doppler broadening parameter, giving $b_5 = 3$ \citep{Draine_1996}. 
We estimate column density using the virial radius of the halo, $N_{\rm H_2} = r_{\rm vir} \times n_{\rm H_2}$, where $n_{H_2}$ is calculated with the incident $J_{0}$ assuming no self-shielding and $r_{\rm vir}$ follows equation (24) of \citet{Barkana_2001}. While the virial radius is conservatively large for this estimate, we adopt it to offset a possibly underestimate of $n_{H_2}$ derived under optically thin conditions. In fact, we find that this crude approximation yields values of self-shielding typically close to $1$ (i.e. no shielding) for the haloes and large incident $J_{\rm LW}$ values explored here. While such a model may significantly underestimate the shielding for small values of $J_{\rm LW}$, we expect that focusing on the haloes with the highest incident radiation will mitigate this issue. As a further test, comparing our model to Fig.~9 of \citet{Kulkarni_2021} with $J_{\rm LW} = 1$ at $z=15$, we find agreement for halo masses below $10^6 M_\odot$, where the higher values of $J_{\rm LW}$ explored here should improve the accuracy of our self-shielding calculation for higher masses. Using this definition for self-shielding, the final LW intensity is then $J_{\rm LW} = f_{\rm shield} J_0$.

\subsection{DCBH candidate selection}

Avoiding gas collapse until the ACT does not guarantee the formation of a SMS. While our MC merger trees have the advantage of efficiently producing the merger history of $10^4$ dark matter haloes, the loss of spatial information requires us to estimate the fraction of DCBH candidates that go on to form SMSs and DCBHs. \citet{Lupi_2021} investigates an over-dense region of haloes, and find that one progenitor of a quasar-hosting halo form a synchronized pair and eventually merge with the quasar host at $z=6$. This synchronised pair forms when a star-forming halo is near ($\leq 1$ kpc) a pristine ACH, illuminating it with a LW flux $\gsim 10^3$, preventing its fragmentation after reaching the atomic cooling stage, bridging the gap between the onset of atomic cooling and SMS formation~\citep{Dijkstra_2008,Visbal_2014b}. \citet{Toyouchi_2023} follow up the MMH and LWH haloes from \citet{Wise_2019}, which were the focus of \citet{Scoggins_2022}, and they find that one of these two haloes go on to form supermassive stars. These investigations set a reasonable lower bound for at least one DCBH candidate per QSO host to eventually form a DCBH. However, the upper bound for the fraction of DCBH candidates that go on to form DCBHs is unclear. 

For the purpose of calculating the OMRL, we here consider two scenarios. In the pessimistic scenario, we assume only the most irradiated halo in each tree, as a proxy for the synchronized pair scenario, goes on to form a DCBH and we discard all other branches for that tree. In an optimistic scenario, we select the 5 most irradiated DCBH sites and assume they go on to form SMSs and DCBHs. This represents ${\sim} 1\%$ of the DCBH candidates in each tree (typically hosting $400$--$1,200$ DCBH candidates, similar to the 1390 pristine QSO progenitors in \citealt{Lupi_2021}). We note that while our model does not explicitly track metal enrichment, which could affect the formation and mass of the SMS, we mitigate this by selecting the most irradiated haloes. We also note that our SMS candidates, by construction, reside in halos in which no progenitor has formed stars (enforced by the criterion that every progenitor has a cooling time longer than the dynamical time).   However, external metal pollution from nearby halos could still reduce somewhat the number of SMS candidates~\citep{Lupi_2021}.

In the optimistic model, it is not clear if the 5 DCBH candidates will merge as their host haloes merge. We simplify accounting for mergers by assuming the two haloes hosting a DCBH  merge the BHs instantly and the resulting black hole remains at the center of the halo. While this oversimplifies black hole mergers, a careful account should be bounded by the optimistic and pessimistic cases excluding accounting for ejection. However, see the Appendix for a discussion of ejection, where we find that it is appropriate to assume the black holes remain in the potential wells of their host halo after a merger.

\subsection{Calculating stellar and black hole mass}
\label{sec:methodsSM}

We assign stellar masses to our haloes following a combination of fitting formulae in two different disjoint halo mass ranges. First, we follow \citet{Behroozi_2019}, which uses a combination of simulation data and observational constraints to fit median stellar mass to halo mass and redshift. Specifically, we adopt the relations in their Appendix~J with constants adopted from their Table~J1. Constants are chosen depending on the following: stellar mass (SM) being true or observed; star-forming vs quenched (SF/Q); satellite or central haloes (Sat/Cen); and including or excluding intrahalo light (IHL). We choose row 15 of the table, corresponding to the true stellar mass for star-forming central and satellite haloes. This only leaves the option to exclude IHL. (SM=True, SF/Q=SF, Sat/Cen=All, IHL=Excl). equation~J1 in \citet{Behroozi_2019} comes from best-fitting the median ratio of stellar mass to peak historical halo mass ($M_{ \rm peak}$), the maximum mass attained over the halo's assembly history. For our MC merger trees which grow monotonically, $M_{\rm peak} = M_{\rm halo}$ at any given snapshot. These formulae were fit and are applied for haloes with mass $10^{10.5} \leq M_{ \rm halo}/{\rm M}_\odot < 10^{15}$, at redshift $z \geq 10$. 

\begin{figure}   
\includegraphics[width=\columnwidth]{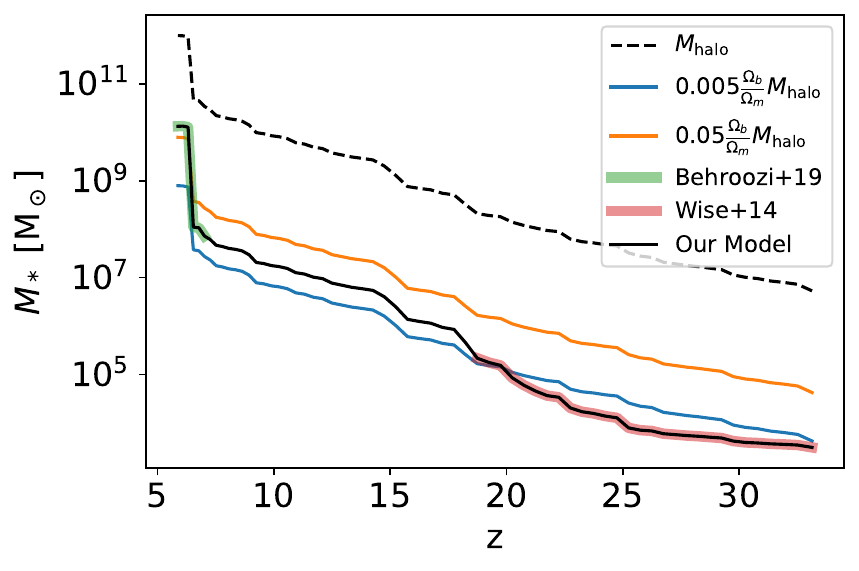}
\caption{We compare several models for calculating the stellar mass. We apply these to a representative dark matter halo branch (shown by the dashed black line), which is the most irradiated DCBH candidate from a randomly selected MC merger tree. The \citet{Behroozi_2019} model is applied within the bounds of the fit, $z\leq 10$ and $M_{\rm halo} \geq 10^{10.5} {\rm M_\odot}$. The stellar mass at the time of DCBH formation and until $M_{\rm halo}$ exceeds $10^{8.5} {\rm M_\odot}$ is calculated using the halo-stellar mass relation from \citet{Wise_2014}, fitting stellar mass to halo mass in a cosmological simulation run until $z=7$ with dark matter haloes $10^{6.5} \leq M_{\rm halo}/{\rm M_\odot} \leq 10^{8.5}$. Between these two fitting formulae, we interpolate in $M_{\rm halo}$-space anchoring the initial mass to the last point provided by \citet{Wise_2014} and the first point provided by \citet{Behroozi_2019}. We also compare this approach to two alternative stellar mass calculations, $M_* = f \frac{\Omega_b}{\Omega_m} M_{\rm halo}$ for $f = 0.05$, $0.005$.
}
    \label{fig:sm_comp}
\end{figure}

The second fitting formula comes from \citet{Wise_2014}, which finds stellar mass and halo mass statistics from a cosmological simulation. In their Table~1, they provide log$(M_{\rm vir})$ and log$(M_*)$ statistics for $6.5 \leq\log(M_{\rm vir}/{\rm M}_\odot) \leq 8.5$ in 0.5 dex bins. We interpolate across log$(M_{\rm vir})$ to derive log$(M_*)$ for a given halo mass and apply this to haloes with $10^{6.5} \leq M_{\rm halo}/{\rm M_\odot} \leq 10^{8.5}$. We note that these statistics are generated from a simulation that ran until $z=7.3$, but we apply them to haloes with redshift $z \geq 6$.
For haloes with a mass between these two bounds, $8.5 \leq\log(M_{\rm vir}/{\rm M}_\odot) \leq 10.5$, we calculate stellar mass by interpolating across halo mass between the smallest mass calculated with \citet{Behroozi_2019} and the largest mass calculated by \citet{Wise_2014}, for every branch. We show an example of our stellar mass calculation in Fig.~\ref{fig:sm_comp}, applied to a randomly selected MC branch. Though the DCBH formation mechanism assumes little to no star formation at the time of forming the SMS and subsequent black hole seed, we follow this stellar mass description which gives generous estimates for the initial stellar mass, making our OMRL calculations conservative.

\begin{figure*}
 \includegraphics[height=5.5cm]{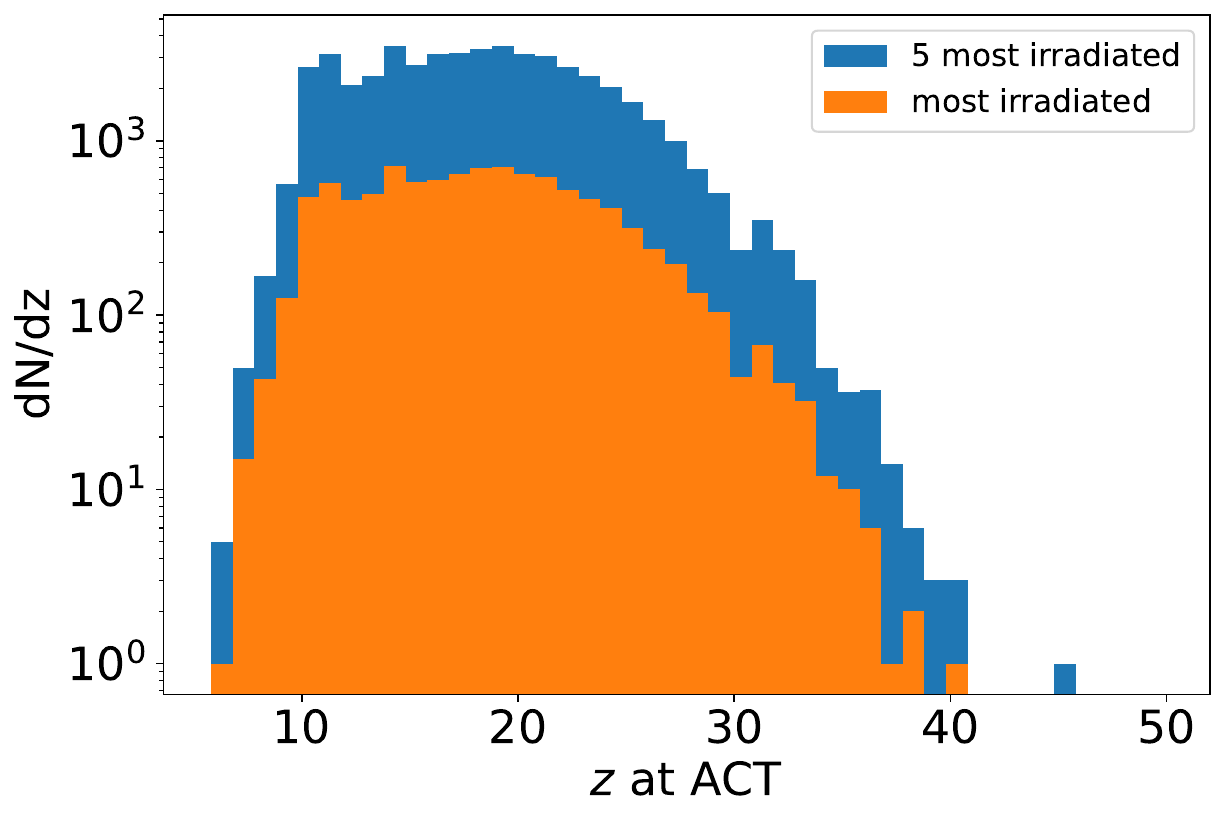}
 \includegraphics[height=5.5cm]{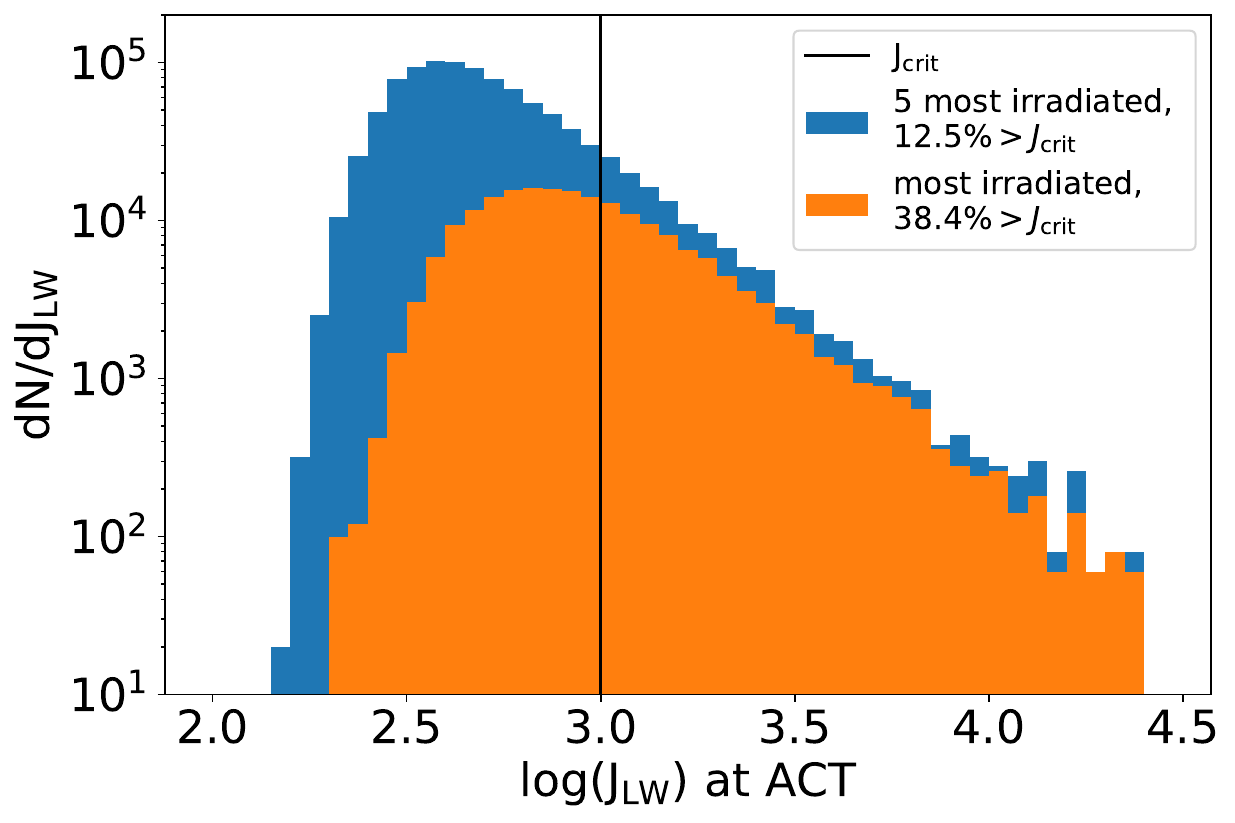}
 \caption{{\it Left:} The redshift distribution of the ACT crossing for our most irradiated DCBH candidates. The most irradiated progenitors, orange, represent the most irradiated haloes at the point of the ACT crossing for each tree. The blue distributions represent the 5 most irradiated DCBH candidates during the ACT crossing. We consider a halo a  DCBH candidate if it reaches this point without collapsing and forming stars before this (we assume this happens if the cooling time exceeds the Hubble time at all snapshots for all progenitors prior to this crossing).
 {\it Right:} Showing the same haloes as the left figure, but plotting the distribution of the Lyman-Werner radiation intensity they experience at ACT crossing, $J_{\rm LW}$, and noting the fraction of ACH sites with $J_{\rm LW} > J_{\rm crit}$.} 
 \label{fig:ach_dist}
\end{figure*}

Black holes are assumed to form shortly after the haloes reach the ACT. Similar to S22, we explore a range of parameters. Initial seed black hole masses in the \texttt{Renaissance} simulation are estimated to fall within the range $10^4  {\rm M_\odot}\leq M_{\rm bh} \leq 10^6 {\rm M_\odot}$ where gravitational collapse to a SMBH is triggered by relativistic instability. We note that a re-simulation of two of the atomic cooling haloes in the Renaissance suite found lower SMS masses of $M{\approx}10^2-10^{4} M_\odot$ \citep{Regan_2020b}, with higher $J_{\rm LW}$ yielding a higher mass.  However, the haloes in this re-simulation experienced a much smaller $J_{\rm LW}$ (${\sim}10 J_{21}$) than we investigate here $({\sim} 10^3 J_{21})$, so we expect our seeds to be much more massive. We estimating the initial black hole mass to be some fraction of the baryonic material, $M_0 = f_{\rm cap}\frac{\Omega_b}{\Omega_m}M_{\rm halo}$, with $f_{\rm cap}  \in \{ 0.1, 0.5\}$. This typically yields black holes with masses $10^4 - 10^5 {\rm M_\odot}$. The growth of these black holes is assumed to follow the Eddington rate
\begin{align}
   &\dot{M}_{\rm bh} = \frac{L_{\rm edd} }{\epsilon c^2} = \frac{4\pi G \mu m_{\rm p} M_{\rm bh}}{\sigma_{\rm T}c \epsilon} = \frac{M_{\rm bh}}{\tau_{\rm fold}}
\end{align}
with speed of light $c$, gravitational constant $G$, mean molecular weight $\mu$ ($\mu \sim 0.6$ for ionised primordial H+He gas), proton mass $m_{\rm p}$, Thomson cross section $\sigma_{\rm T}$,  and radiative efficiency $\epsilon$. This leads to a black hole mass given by $M_{\rm bh}(t) = M_0\exp(t/\tau_{\rm fold})$ with e-folding time $\tau_{\rm fold} =  (\sigma_{\rm T} c \epsilon)/(4\pi \mu G m_{\rm p}) \approx  $ 450$\epsilon$ Myr. Assuming efficiency $\epsilon \approx 0.1$, we consider $\tau_{\rm fold} \in \{40, 80\}$ myr. We additionally quench black hole growth when the mass of the black hole exceeds a prescribed fraction of the baryonic matter in the halo, capping $M_{\rm bh} \leq f_{\rm cap} M_{\rm halo} \Omega_{\rm b}/\Omega_{\rm m}$. To summarise, our simple model governs black hole growth through $f_{\rm cap}, \tau_{\rm fold}$, $M_{\rm halo}$, and $M_0$ (which is determined by $f_{\rm cap}$ and $M_{\rm halo}$).

We start the growth of our black holes immediately after formation. While stellar feedback could initially stunt black hole accretion in most ACHs, recent work has suggested that black holes born in biased progenitor halos that end up in very massive haloes (such as the $M{\sim} 10^{12} M_\odot$ host explored in this work at $z=6$) do not experience significant stunting (e.g. see the comparisons between the left and right panels in Fig.~12 in \citealt{Inayoshi_2020}).

\subsection{Calculating the over-massive relation lifetime (OMRL)}

We define the lifetime for a SMBH to satisfy an unusual mass ratio as $\tau_{\rm OMRL} = t_f - t_0$ where $t_0$ is the time when the black hole is formed and $t_f$ is the time when the black hole first crosses the minimum threshold for $M_{\rm bh}/M_*$, typically chosen to be unity but other values are explored below. This value gives a generous threshold where the mass relation is unambiguously above the light seed formation pathway (${\sim}10^{-2}$), the high-$z$ QSO mass relation (${\sim}10^{-2}$) and the local SMBH relation (${\sim}10^{-3}$).

\section{Results}
\label{sec:results}

\subsection{DCBH candidates and halo evolution}

\begin{figure}
 \includegraphics[width=\columnwidth]{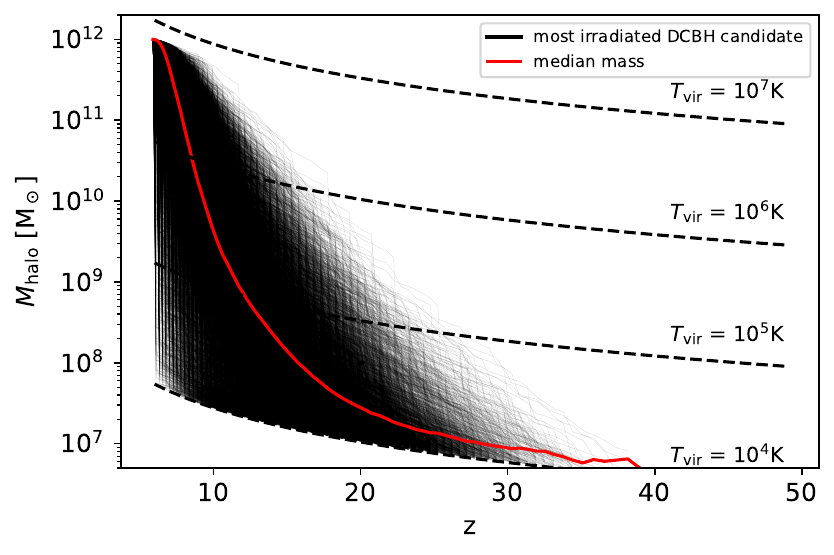}
 \caption{The evolution of the most Lyman-Werner irradiated DCBH candidate in each MC merger tree, beginning from the time when the halo crosses the atomic cooling threshold (ACT). The subsequent median mass of these haloes is shown in red. Dashed lines show the virial temperature, and we assume crossing the ACT happens when halo virial temperatures reach $10^4$K. The curves near the bottom left represent small haloes that merge with the $10^{12} {\rm M_\odot}$ halo near redshift $z{\sim} 6$.}
\label{fig:mass_vs_redshift}
\end{figure}

In Fig.~\ref{fig:ach_dist}, we show the redshift distribution and the $J_{\rm LW}$ distribution of our DCBH candidates at the time of crossing the ACT for the most irradiated haloes (orange, the pessimistic case) and the 5 most irradiated haloes (blue, the optimistic case) from each MC merger tree. Following the method laid out in \S~\ref{sec:methods}, these DCBH candidates are haloes that reach $T_{\rm vir} = 10^4 $K while satisfying the no-cooling condition $t_{\rm cool} > t_{\rm hub}$ at all snapshots for every progenitor. These extremely irradiated haloes cross this threshold at somewhat larger redshifts than in previous works, where the time of ACT crossing is typically dominated by haloes at redshift $z\sim 10 - 15$. (e.g. see Fig.~2 in \citealt{Lupi_2021}). Our distribution is dominated by haloes crossing closer to $z\sim 15-20$. 

There are likely two reasons for this. The first is that our selection of the most irradiated haloes with $T_{\rm vir} = 10^4 {\rm K}$ prefers lower mass (higher redshift) due to $\overline{J}_{\rm LW}$ tending to grow with redshift along the $10^4 {\rm K}$ contour, until $z {\sim} 30$. (see Fig. 2 of  \citealt{Li_2021}, where they explore the evolution of the primary progenitors of MC merger trees and find that the median $J_{\rm LW}$ tends to grow up to ${\sim} 10^3$ at redshift $z=30$, then sharply declines at higher redshift). This non-monotonic behavior can be explained by the onset of star formation, which causes the initial increase in $\overline{J}_{\rm LW}$, eventually being offset by the merger of star-hosting haloes. These mergers cause the average distance between active regions to begin to grow and outpace the contribution from star formation, resulting in a steady decline of  $\overline{J}_{\rm LW}$.
The second reason that our redshift distribution is higher than in previous work is due to the nature of MC merger trees. Other works which investigate ACHs and the redshift of the ACT crossing may compare haloes in a comoving volume, but do not guarantee that they merge into the SMBH's halo near redshift $z=6$, whereas our MC merger trees focus on haloes in extremely biased dense regions which are guaranteed to end up in the $10^{12} {\rm M_\odot}$ halo at redshift $z=6$ by construction. This biases our selection to the slightly more massive progenitors which tend to cross the ACT at higher redshifts. Given this, the most irradiated haloes at the time of ACT crossing represent the outliers, and the majority of the DCBH candidates cross this threshold at lower redshifts.

\begin{figure*}
 \includegraphics[width=\columnwidth]{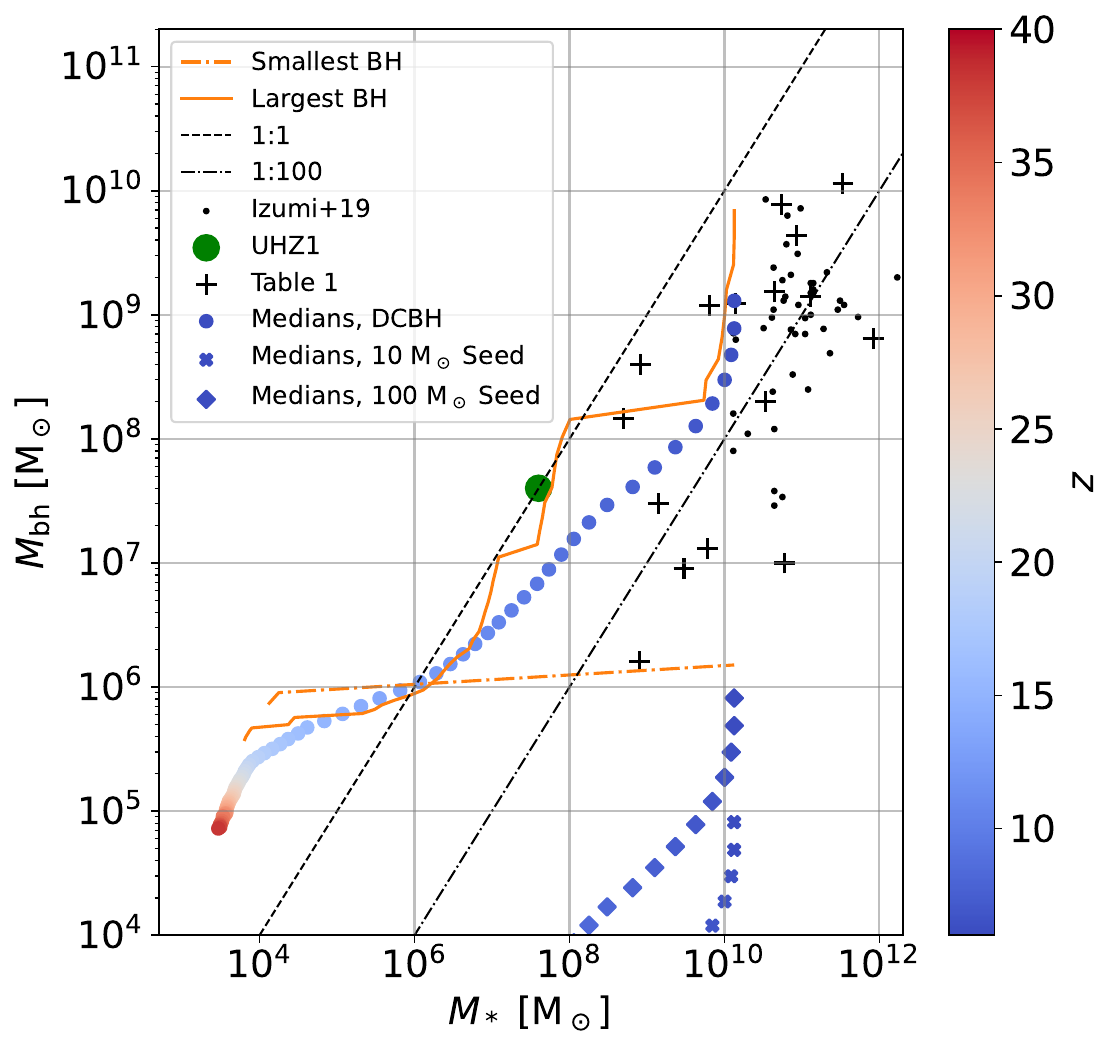}
  \includegraphics[width=\columnwidth]{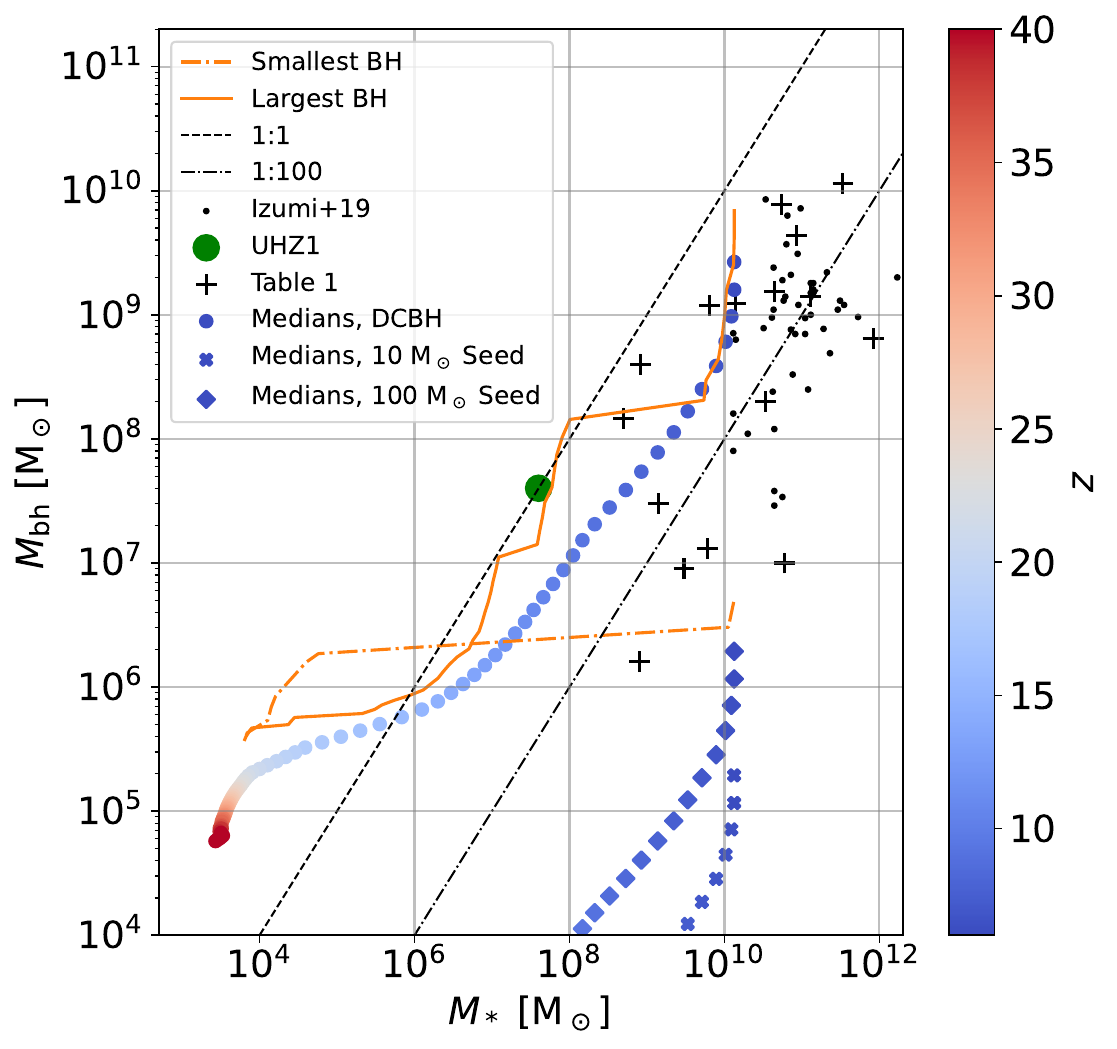}
  \includegraphics[width=\columnwidth]
  {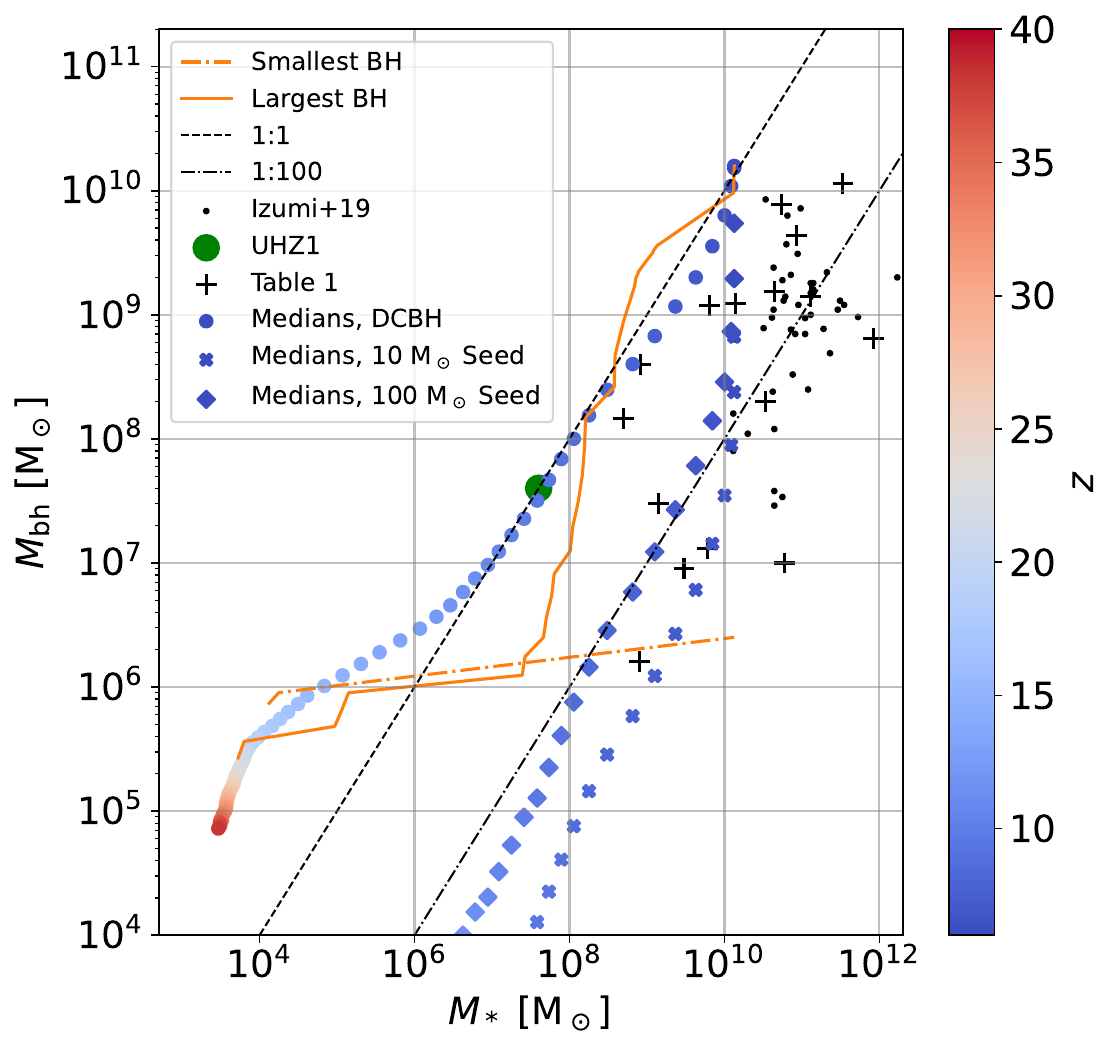}
 \includegraphics[width=\columnwidth]{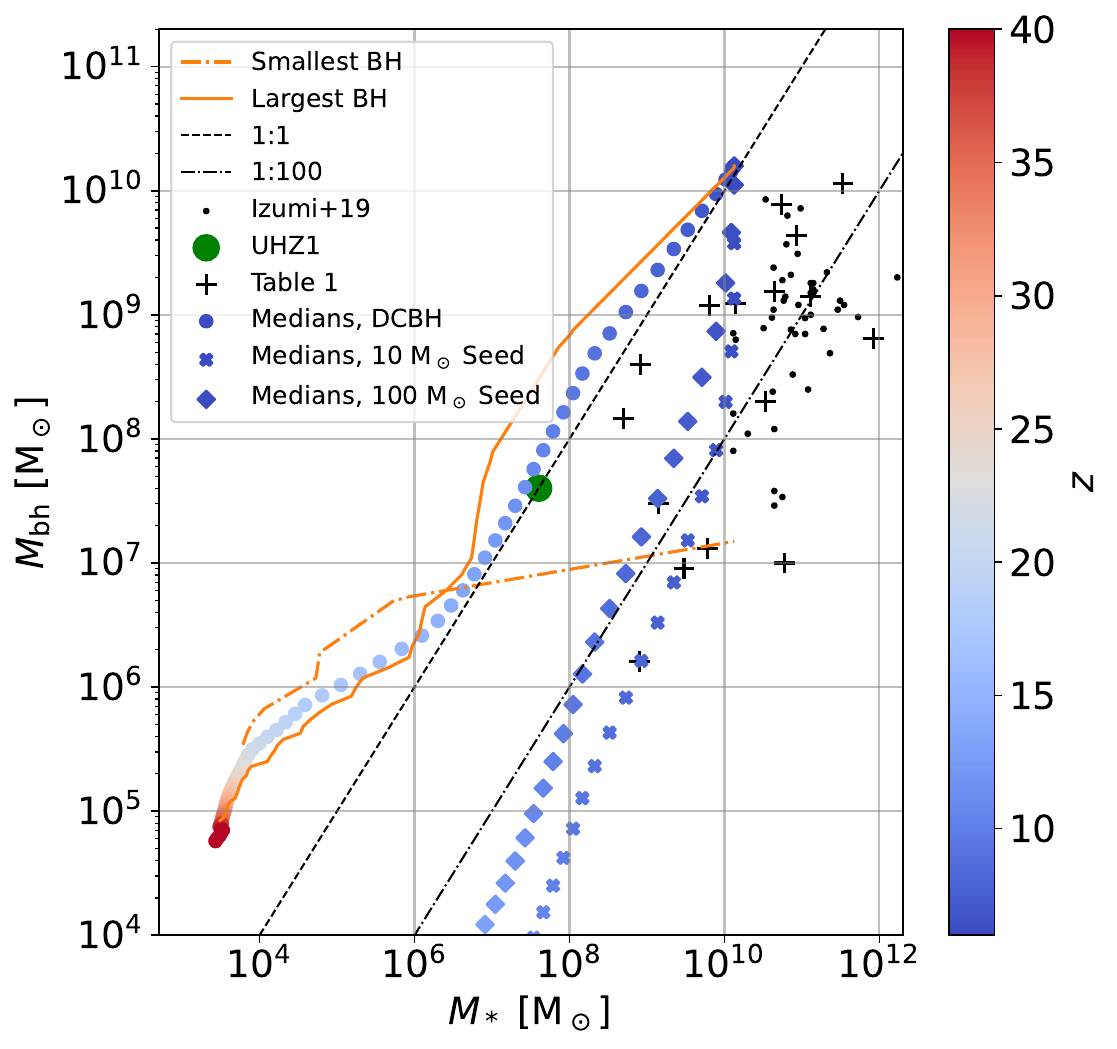}
 \caption{The co-evolution diagram comparing black hole and stellar mass. Orange shows black holes with the most (solid) and least (dashed) massive final mass, along with the median black hole and stellar mass for our DCBHs (blue circles). We compare our DCBH evolution to their light-seed counterparts, with growth parameters being the same but starting with $10 {\rm M_\odot}$ and $100 {\rm M_\odot}$ seeds. Left represents the pessimistic case where only the most irradiated halo of each tree forms a DCBH and the right shows the optimistic case where the 5 most irradiated haloes form DCBH candidate sites from each tree form a  DCBH and eventually merge. For $\tau_{\rm fold} = 80$ (top), the black holes rarely reach the cap we imposed by the fraction $f_{\rm cap}=0.1$ of the total baryonic mass in the halo, and the discrepancy in mass between the three seeds is roughly fixed over different values of $M_*$. With more efficient growth, $\tau_{\rm fold} = 40$ (bottom), the final mass is roughly independent of initial seed mass, as the growth is limited by the cap. Grey points show the high-$z$ quasar samples compiled by \citealt{Izumi_2019}, with stellar mass calculated from [C II]-based dynamical mass conversions calibrated in low redshift galaxies (\citealt{Tacconi_2018}; see also \citealt{Hu_2022b}). We also plot the recent {\it JWST} observations compiled in Table~\ref{table1} (crosses).}
 \label{fig:coevolution}
\end{figure*}

We also show the value of $J_{\rm LW}$ at the time of the ACT crossing in Fig.~\ref{fig:ach_dist}. While previous work has found that avoiding star formation and achieving DCBH candidacy requires $J_{\rm LW} \geq J_{\rm crit} = 10^3$, most of our DCBH haloes do not experience these levels of radiation, as dynamical heating from rapid mergers contributes to offsetting most ${\rm H_2}$ cooling, preventing fragmentation and star formation prior to the ACT. For the pessimistic case, $38.4\%$ of our $10,000$ ACHs experience $J_{\rm LW} \geq J_{\rm crit}$. For the optimistic case, $12.5\%$ of our $50,000$ ACHs experience $J_{\rm LW} \geq J_{\rm crit}$.

In Fig.~\ref{fig:mass_vs_redshift} we show the evolution of the most irradiated DCBH candidate from each MC merger tree, as well as the median mass of these haloes above $T_{\rm vir} = 10^4{\rm K}$ for each snapshot. We also compare the co-evolution of black holes and the stellar mass of their hosts in Fig.~\ref{fig:coevolution}. We show the evolution of the most and least massive black hole at $z=6$, as well as the median black hole and stellar mass for each snapshot. Left panels show the pessimistic case and right panels show the optimistic case. All panels show black hole growth with $f_{\rm cap} = 0.1$, though top panels show $\tau_{\rm fold} = 80$ myr and the bottom panels show $\tau_{\rm fold} = 40$. For reference, we show the high-$z$ quasar samples compiled by \citet{Izumi_2019}. We also show the $M_{\rm bh}/M_*$ ratio of 1:1  (the ratio we typically use in most of our OMRL evaluations in the next section) along with a $1:100$, the standard ratio for the Pop III formation pathway and most of the observed SMBHs at high redshift. We compare the evolution of our DCBHs to their light-seed counterparts, using the same model for growth, but with an initial mass of $10 {\rm M}_\odot$ and $100 {\rm M}_\odot$. We also compare these results to recent {\it JWST} observations, with their $M_{\rm bh}$ and $M_*$ compiled in Table~\ref{table1}.

The largest black hole at $z=6$ is similar for all panels ($\sim$$10^{10} {\rm M_\odot}$) with a mass ratio of nearly 1:1. The smallest black hole varies by almost an order of magnitude for different models of BH growth, being as small as $10^6 {\rm M}_\odot$ and up to  $10^7 {\rm M}_\odot$, with a mass ratio well below $10^{-2}$. The smallest black holes represent the late-forming DCBHs which then quickly merge with the $10^{12} {\rm M}_\odot$ halo at $z=6$, leaving little time for BH growth.

The median black hole mass is larger in the optimistic cases than in the pessimistic cases for any given stellar mass above $M_* > 10^8 {\rm M_\odot}$, but the pessimistic cases have larger black holes below this stellar mass. This is likely due to the most irradiated haloes typically being more massive (as $\overline{J}_{\rm LW}$ increases with mass) and initially experiencing a smaller halo growth, allowing the hosted black hole to grow faster relative to the surrounding stellar mass. In both cases, the black holes initially start with a ratio of ${\sim} 10$, then grow slightly, before reaching 1 near $M_* = 10^6 {\rm M_\odot}$. This initial ratio of our black holes is indicative of the stellar mass calculation over-predicting the initial stellar mass, where DCBHs typically have ratios closer to $10^3$.

Comparing the light seed and heavy seed models in Fig.~\ref{fig:coevolution}, we find that the final mass varies dramatically depending on the chosen $\tau_{\rm fold}$. We also note that the influence of mergers is negligible on final median mass of our black holes (comparing the left panels to the right panels). With extremely aggressive black hole growth, ($\tau_{\rm fold} = 40$, bottom panels), light seeds formed in these ACHs can account for the SMBHs observed at high redshift, but even in this case, the mass relation at higher redshift ($z\geq 10$) is typically below $10^{-2}$. If we compare the light and heavy seed models in this figure to the recent high-redshift low-mass SMBH observations, we find that almost every observation is more consistent with the light seed model, with the exception of UHZ1. See \S~\ref{sec:discussion} for further discussion of these observations.

\begin{figure*}
 \includegraphics[width=\columnwidth]{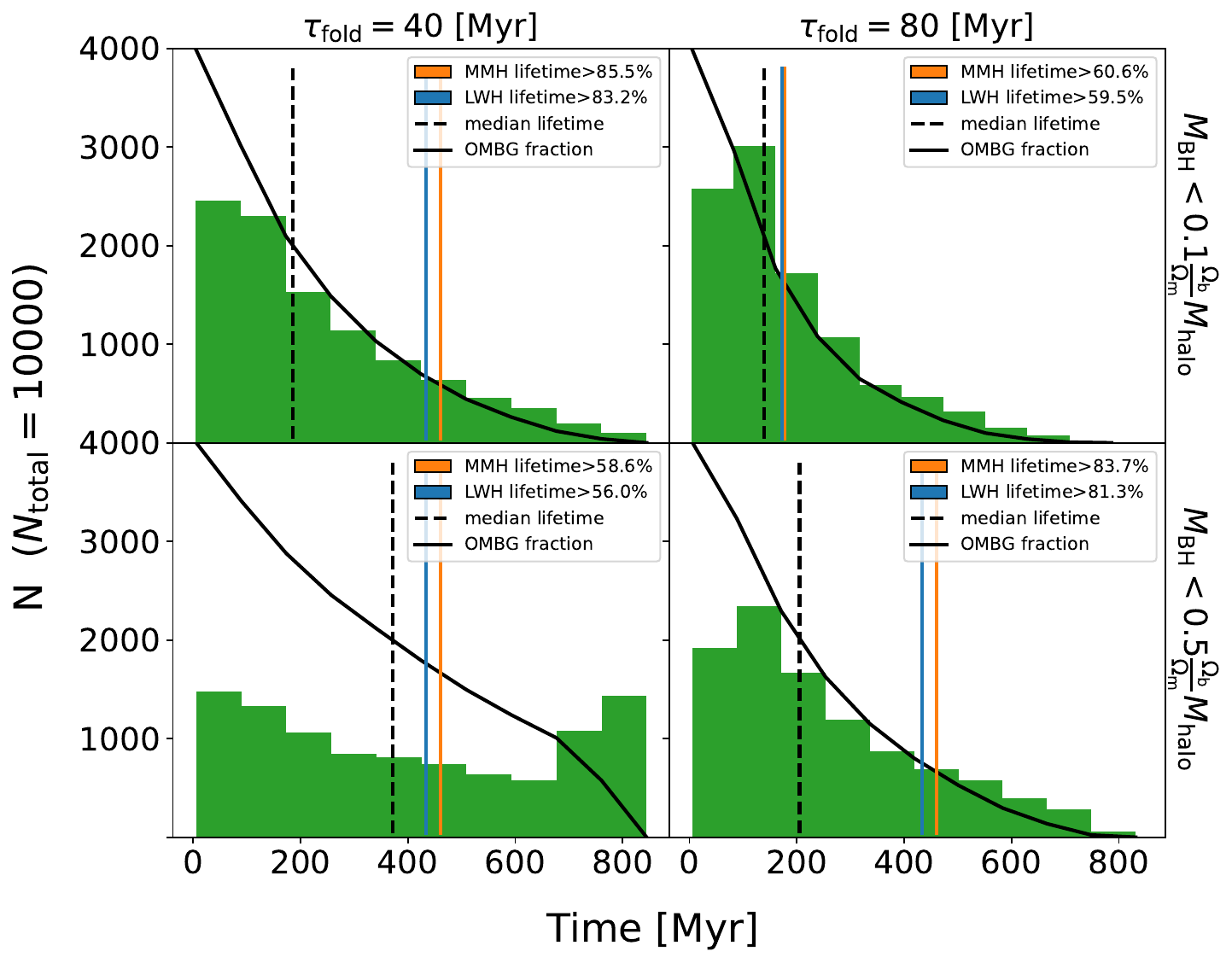}
 \includegraphics[width=\columnwidth]{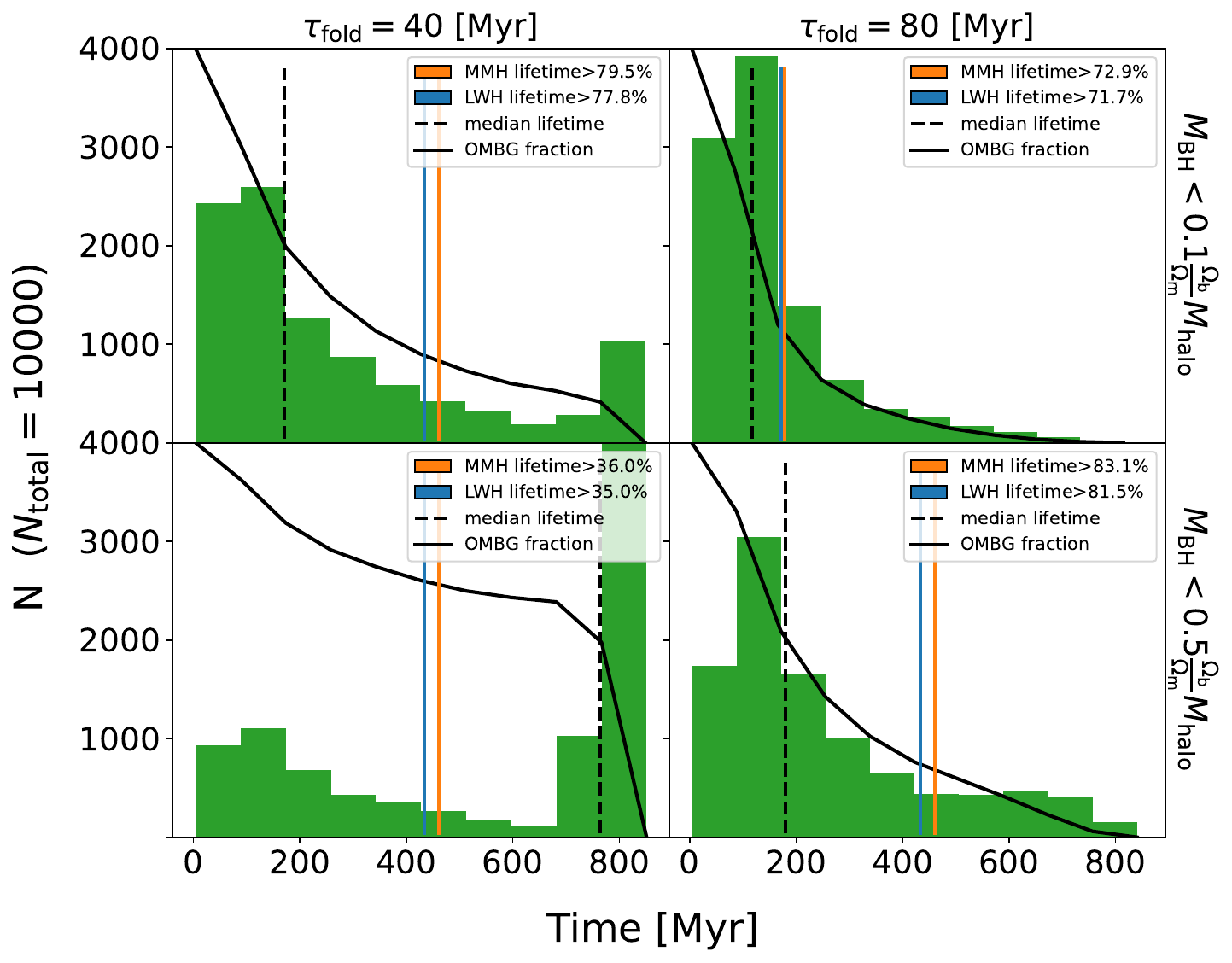}
  \includegraphics[width=\columnwidth]{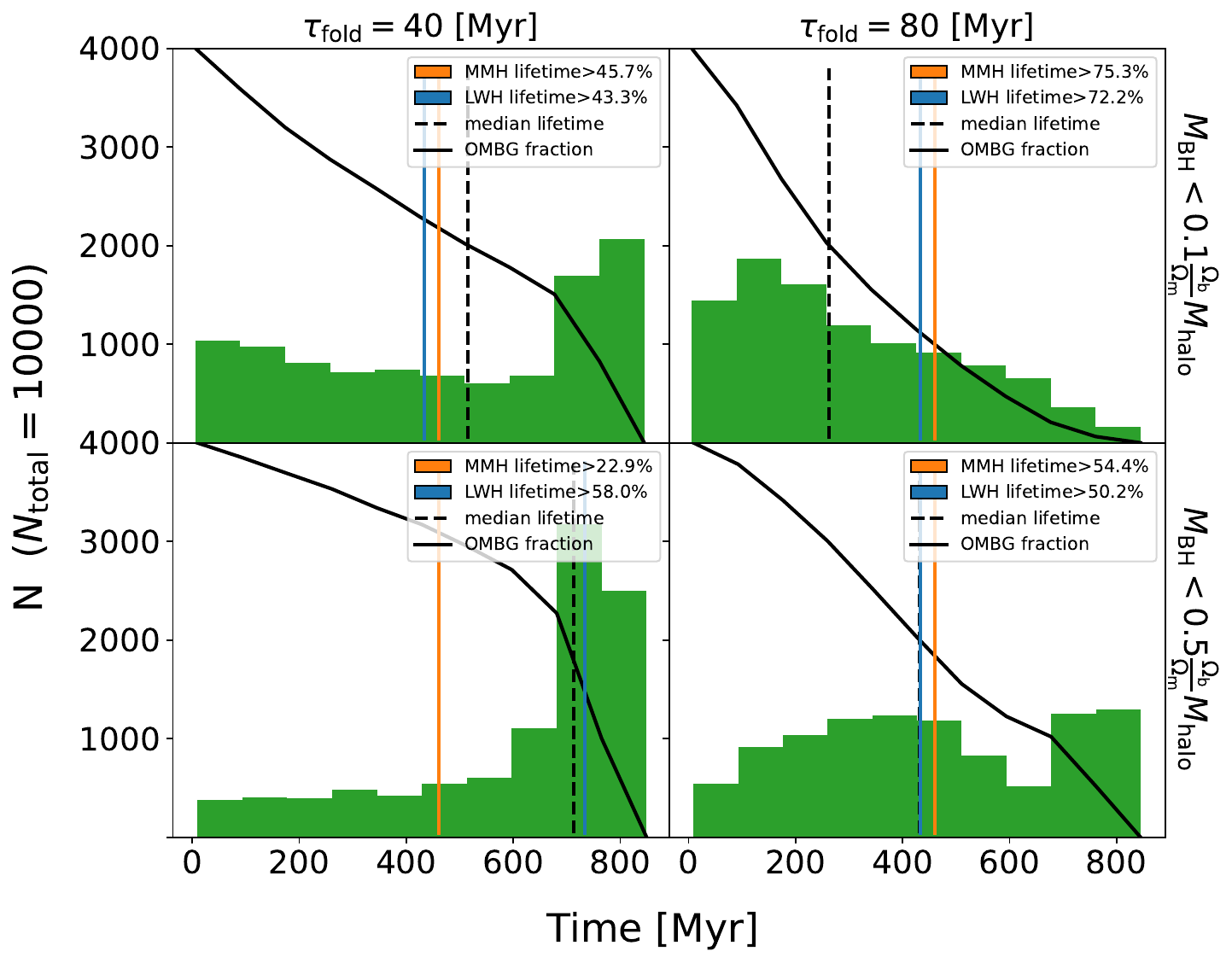}
 \includegraphics[width=\columnwidth]{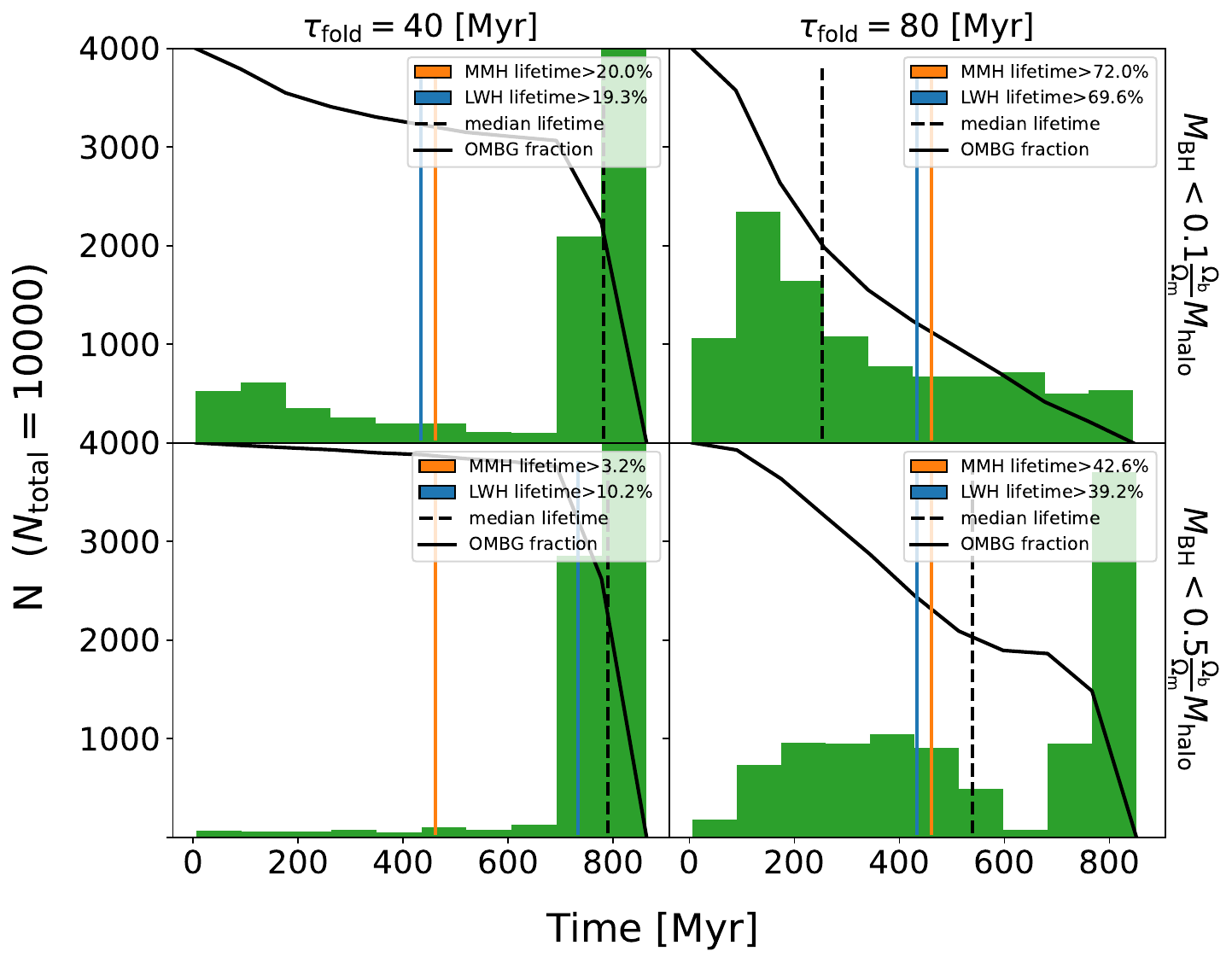}
 \caption{The over-massive relation lifetime (OMRL) distribution for our DCBH candidate haloes. The OMRL is calculated using the difference in time between the assembly of the black hole (assumed to happen almost immediately after crossing the ACT) and the first instance when $M_{\rm bh}/M_* < 1$ (top) and $M_{\rm bh}/M_* < 0.1$ (bottom). Left shows the case where only the most irradiated DCBH candidate forms a massive seed. Right shows a more optimistic assumption for growth, where the 5 most irradiated DCBH in each tree form a massive seed, and the black holes in each tree merge before $z{=}6$, though we only plot the OMRL of the earliest DCBH candidate halo. Dashed vertical lines show the median lifetime for each distribution and the solid black curves show the fraction of OMBGs which still hold an outstanding relation, ranging from 1 in the top left and ending at 0 in the bottom right of each panel. We compare these OMRLs to the MMH (shown in orange) and LWH (shown in blue) haloes explored in \citet{Wise_2019} and \citet{Scoggins_2022}. These OMBG candidates are hosted by haloes that experience slow growth until merging with a much more massive halo at redshift $z{=}8$, making them less sensitive to growth parameters.}
 \label{fig:lifetime_dist}
\end{figure*}

\subsection{The over-massive relation lifetimes of the DCBHs}

In Fig.~\ref{fig:lifetime_dist}, we calculate distribution of the
over-massive relation lifetimes (OMRLs) of the DCBHs, which, as defined above, is the total time elapsed from black hole formation until the $M_{\rm bh}/M_*$ relation falls below a fixed ratio $M_{\rm bh}/M_* \leq 1$ (top) and $M_{\rm bh}/M_* \leq 0.1$ (bottom). We compare the OMRL distributions for several black hole growth parameters, with $\tau_{\rm fold} \epsilon \{40, 80\}$ Myr and $f_{\rm cap} \epsilon \{0.1, 0.5\}$, for both the pessimistic (left) and optimistic (right) case. We also compute the fraction of the DCBHs which have maintained their over-massive signature for a given duration (i.e. $1-$CDF, where CDF is the cumulative distribution function), shown in black.  For models with the most aggressive BH growth (the bottom left panels), most lifetimes exceed $600$ Myr. For the least aggressive BH growth (the top right panels), the lifetimes are much shorter, where the median is usually ${\sim} 200$ Myr. 

Comparing these distributions to the most massive halo (MMH) and most Lyman-Werner irradiated halo (LWH) from S22, these target haloes are not necessarily outliers, though we note that their OMRL is not sensitive to the growth parameters. This is caused by the growth of the MMH and LWH haloes being relatively modest until a merger with a much larger halo near redshift $z{=}8$, meaning the MMH and LWH have a well establish OMBG relation for most growth parameters until this merger wipes out the OMBG property after ${\sim} 400$ Myr.

\begin{figure*}
   \includegraphics[width=1\columnwidth]{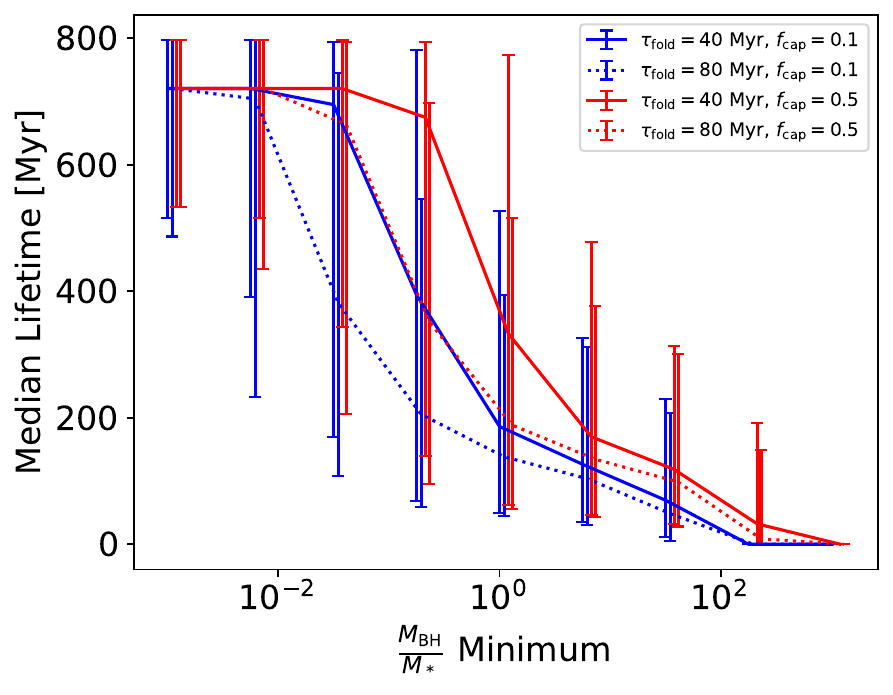}
   \includegraphics[width=1\columnwidth]{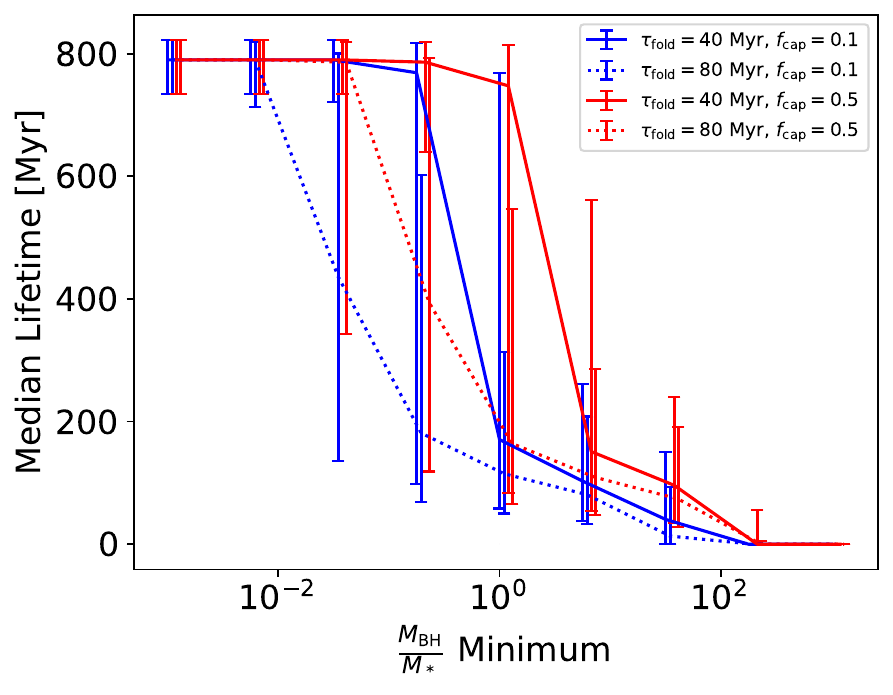}
 \caption{The median over-massive relation lifetime (OMRL) {\it vs.} the minimum ratio which determines the OMRL, for a pessimistic case, assuming only the most irradiated DCBH site forms a SMS and BH seed (left), and an optimistic case assuming the 5 most irradiated DCBH sites form BH seeds (right) and with 80\% error bars. For a minimum ratio of 0.1, more than half of the sites in both cases live through $z\leq 10$, with the optimistic case yielding an even higher fraction.}
 \label{fig:lifetime_vs_ratio}
\end{figure*}

The median OMRL in Fig.~\ref{fig:lifetime_dist} is calculated with a minimum ratio of $\frac{M_{\rm bh}}{M_*}=1$ (top) and  $\frac{M_{\rm bh}}{M_*}=0.1$ (bottom), but we explore the effect of varying this ratio in Fig.~\ref{fig:lifetime_vs_ratio}. We plot the median OMRL against the minimum ratio $\frac{M_{\rm bh}}{M_*}$, with error bars showing $10$th (bottom) and $90$th percentile (top) of the OMRLs. As usual, the left shows the pessimistic case and the right shows the optimistic case. We find that with a minimum ratio similar to local values of $10^{-3}$, most of the black holes have a OMRL greater than 600 Myr. At the other extreme with a minimum ratio of $10^3$, nearly 100\% of the black holes drop below this immediately. This is conservative though, as our initial stellar mass calculations are generous given the DCBH scenario meaning our initial $M_{\rm bh}/M_*$ ratios are also conservative. 

With a minimum ratio of $10^{-1}$, an order of magnitude above the ratio for SMBHs at high redshift, the median values vary from 300 to 700 Myr depending on the model for black hole growth. This means that some of these black holes will be detectable into a redshift just beyond the redshift of the observed quasars near $z=6$, with most observable at even higher redshifts. This means the heavy seed mechanism should be distinguishable from other formation pathways.

\subsection{Number density of OMBGs}

\begin{figure*}
\includegraphics[width=\columnwidth]{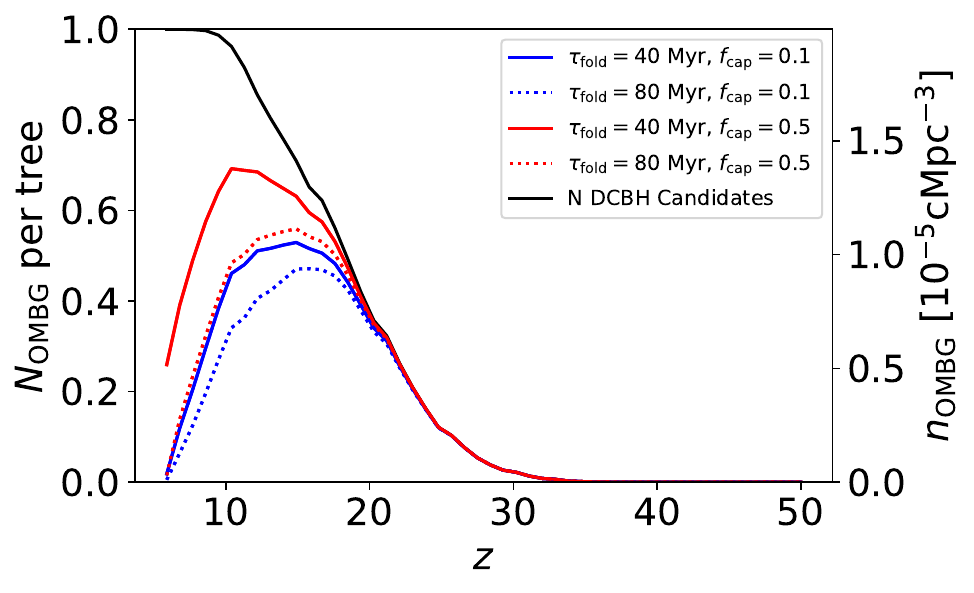}
\includegraphics[width=0.95\columnwidth]{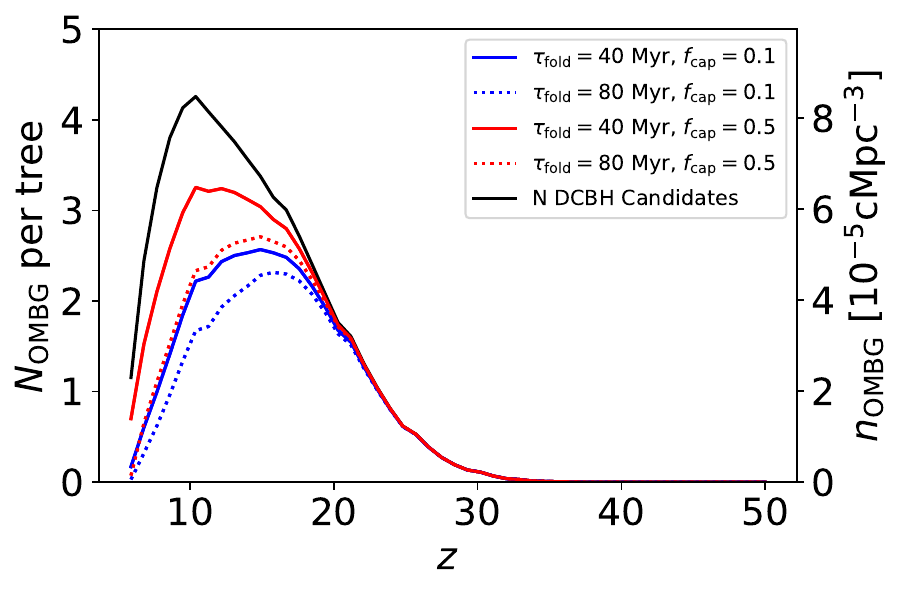}
\includegraphics[width=\columnwidth]{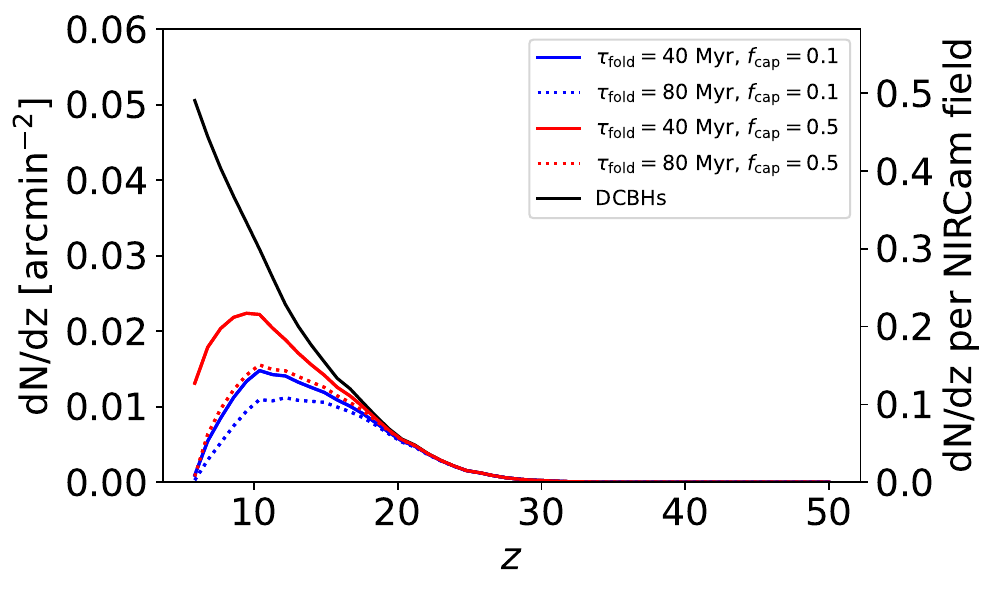}
\includegraphics[width=\columnwidth]{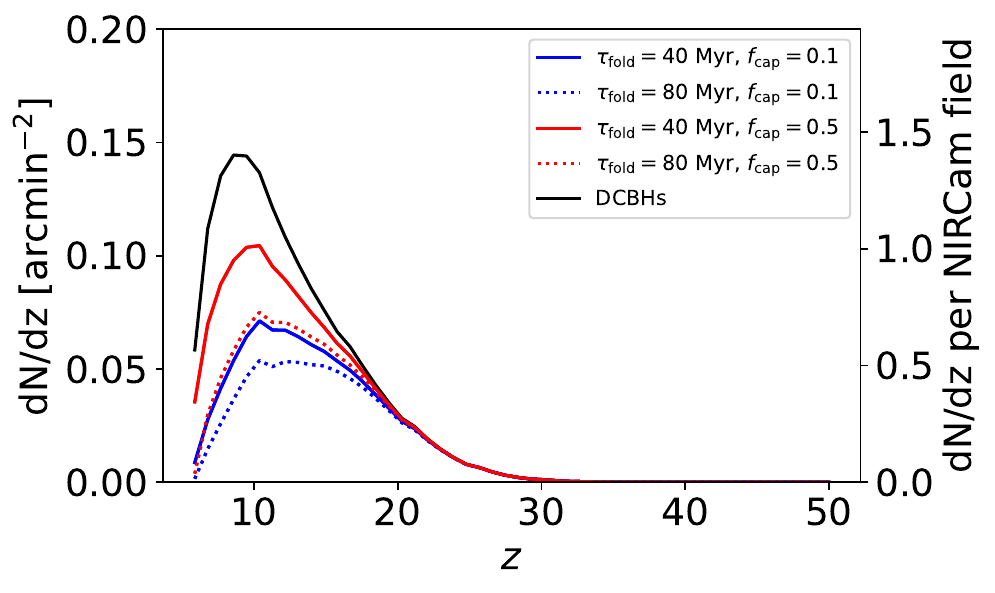}
\caption{{\it Top:} The average number OMBGs per tree (i.e. per $10^{12}~{\rm M_\odot}$ halo) with $M_{\rm bh}/M_* > 1$, evaluated at each redshift. We compare different growth models, varying the e-folding time $\tau_{\rm fold}$ and black hole mass cap $f_{\rm cap}$. The black line shows the total number of DCBHs, regardless of the relation between black hole mass and stellar mass. The right vertical axis labels show the corresponding number density, given that the abundance of haloes with mass $11.5 \leq\log(M_{\rm halo}/{\rm M_\odot}) \leq 12.5$ at redshift $z =6$ is $n_{\rm 1e12} = 2\times 10^{-5}$ cMpc$^{-3}$.
{\it Bottom:} The total number of outstanding ($M_{\rm bh}/M_* > 1$) haloes shown per unit redshift per square arcmin. The black lines show the total number of DCBH candidates. The vertical right axis labels give the expected number of objects per unit redshift per {\it JWST} NIRCam field.
The left columns again show the pessimistic case, and the right shows the optimistic case.}
    \label{fig:n_vs_z}
\end{figure*}

Given that the OMRLs of the DCBHs are maintained into a redshift detectable by {\it JWST} and X-ray surveys (see \S~\ref{sec:discussion} for a discussion of detecting this mass relation), we are motivated to calculate their expected number density. First, we calculate $\overline{N}_r(z)$, the average number of haloes that have a mass ratio above $r$ at redshift $z$, by averaging the total number of haloes with an outstanding relation across all 10,000 trees for each snapshot. The results are shown in the top panels of Fig.~\ref{fig:n_vs_z}, varying the parameters for BH growth, with the total number of DCBH sites shown in black. $\overline{N}_r(z)$ represents the expected number of outstanding haloes for every ${\sim} 10^{12} {\rm M_\odot}$ halo near redshift $z=6$. The results are very sensitive to the number of DCBH candidates which actually go on to form DCBHs. The top left, showing the pessimistic case of one DCBH per tree, sets a lower bound for the expected number of outstanding DCBH sites per $\sim 10^{12} {\rm M_\odot}$ halo as a function of redshift. At redshift $z > 20$, less than $1/3$ of DCBHs have formed. DCBH formation is complete near redshift $z = 10$ when the total number of DCBH candidates approaches 1. The expected number of OMBGs varies for each growth parameter but tends to peak near redshift $z=12$. The results for the top right panel (the optimistic case, assuming 5 DCBHs per tree) are similar in shape to the top left panel, though larger in magnitude. The number of outstanding sites again peaks near redshift $z=12$ for every model for growth. The total number of DCBH candidates does not flat-line, instead peaking near redshift $z=12$, with  $\overline{N}_r(12){\sim} 4.2$, then approaching $1$ as the DCBHs merge.

The comoving number density of haloes with mass $11.5 \leq\log(M_{\rm halo}/{\rm M_\odot}) \leq 12.5$ at redshift $z =6$ is $n_{\rm 1e12} \approx 2\times 10^{-5}$ cMpc$^{-3}$ (calculated using the halo mass function in \citealt{HMF_2013}). We approximate the DCBH results from our MC merger trees as being representative of haloes in this mass range and use this number density to determine the expected number density for outstanding DCBHs. The results of this conversion are shown by the labels on the right axis of the top panels in Fig.\ref{fig:n_vs_z}. We check the consistency of our DCBH number density against the results from \citet{Regan_2020}, where they calculate a DCBH seed number density of 0.26 ${\rm cMpc}^3$ in the \texttt{Renaissance} simulation. Accounting for the rarity of the simulated over-density, they conclude that the global number density should be 3 to 4 orders of magnitude smaller. This results in a global DCBH seed density of $\sim2.6\times 10^{-5}$--$2.6\times 10^{-4}$ ${\rm cMpc}^{-3}$. This lower bound is greater than the number density predicted from our pessimistic case (which predicts a maximum number density of ${\sim} 2\times 10^{-5}$ ${\rm cMpc}^{-3}$), suggesting that our pessimistic case is extremely conservative. The results from our optimistic case, with a peak number density of $8\times 10^{-5}$ ${\rm cMpc}^{-3}$, are in better agreement with the results from \citet{Regan_2020}.

Combining $n_{\rm 1e12}$  with the physical volume per unit redshift per unit solid angle, $\frac{dV}{d\Omega dz} = d_{\rm A}^2(z)c \frac{dt}{dz}$ where $\frac{dt}{dz} = \frac{1}{H(z)(1+z)}$, $d_{\rm A}(z) = \frac{d(z)}{1+z}$ is the angular diameter distance, and $d(z)$ is the comoving distance,  then the number of outstanding DCBH sites per unit redshift per solid angle is given by
\begin{align}
    \frac{dN}{dzd\Omega}(z) &=  n_{\rm DCBH}(z) \frac{dV}{d\Omega dz} (1+z)^3\\
    &= c \overline{N}_r(z) \ n_{\rm 1e12} \frac{d(z)^2}{H(z)}
\end{align} where $n_{\rm DCBH}(z) = \overline{N}_r(z) n_{\rm 1e12}$ is the outstanding DCBH comoving number density. The results are shown in the bottom panels of Fig.~\ref{fig:n_vs_z}. Again, the optimistic and pessimistic cases are similar in  shape for the outstanding DCBHs but differ in magnitude. Within the redshift range $z=6-15$, in the pessimistic case, we expect there to be $\gtrsim 0.01$ DCBHs arcmin$^{-2}$ dz$^{-1}$, or roughly $10^6$ dz$^{-1}$ on the sky in total per unit redshift. With a {\it JWST} NIRCam field of $9.7$ arcmin$^2$, we expect up to $0.1$ objects per field per unit redshift. For the optimistic case, we expect roughly ${\sim} 5 \times 10^6 $ dz$^{-1}$, up to $1$ object per {\it JWST} NIRCam field per unit redshift.

\section{Discussion}
\label{sec:discussion}

While this work has focused on MC trees which evolve into a $10^{12} \ {\rm M_\odot}$ halo at redshift $z{=}6$, SMBH host haloes near this redshift can be somewhat larger. \citet{Arita_2023} estimates the masses of 107 quasar hosts at redshift $z\sim 6$ and find them to be ${\sim} 7 \times 10^{12} {\rm M_\odot}$ by the projected correlation function, or ${\sim}7$ times larger than the haloes explored in this work. Larger haloes would be composed of progenitors that experience more frequent mergers or mergers with larger haloes, leading to increased dynamical heating, and likely more DCBH candidates. Being more massive on average, these DCBH candidates could also cross the ACT and form SMSs/black holes at earlier times, resulting in more massive black holes at each redshift. However, the stellar mass would also be larger, so we expect our OMRLs calculated using $10^{12} {\rm M_\odot}$ haloes to be comparable.

Comparing our results to a similar exploration in \citet{Visbal_2018}, where they analyzed a 20 comoving Mpc box, starting at $z=10$, and tracked the evolution of the $M_{\rm bh}/M_*$ relation in ACHs within this volume. They also find that these sites have outstanding relations, though their outstanding relations last $\sim 100$ Myr. We can attribute these differences to two effects: (1) we focus on the haloes that end up in a $10^{12} {\rm M_\odot}$ halo and (2) we consider the evolution prior to the ACT, filtering out haloes that would have experienced star formation. These effects favor more massive, rapidly merging, higher redshift ACHs which would lead to a longer OMRL. The contrast between these two works highlights the idea that forming a DCBH earlier and in an over-dense region (such as the haloes we have explored which merge with a $10^{12} {\rm M_\odot}$ halo), increases the OMRL (see also \citealt{Lupi_2021}).

\subsection{Searching for OMBGs}

Several recent works have focused on detecting and measuring the properties of high redshift SMBHs on the low-mass end, or to image their hosts' stellar light with {\it JWST} (e.g. \citealt{Bezanson_2022, Maiolino_2023, Maiolino_2023b, Kocevski_2023, Larson_2023, Goulding_2023, Natarajan_2023, Whalen_2023, Lambrides_2023, Nabizadeh_2023, Furtak_2023, Yue_2023, Pacucci_2023, Kokorev_2023, Harikane_2023, Ubler_2023, Barro_2023, Matthee_2023, Stone_2023,Ding_2023, Juodzbalis_2024,Kovacs_2024}). In this section, we briefly discuss some of these observations and note these SMBHs approach the mass range where they can be probed by the $M_{\rm bh}/M_*$ relation. We compile these low-mass SMBHs in Table~\ref{table1}. Establishing the SMBH's location on the $M_{\rm bh}/M_*$ relation will help distinguish between heavy and light seeds.

\begin{table*}
\centering
\begin{tabular}{||c | c | c | c | c ||} 
 \hline
 Source/ID & z & $M_{\rm bh}/{\rm M_\odot}$ & $M_*/{\rm M_\odot}$ &  Reference \\ [0.5ex] 
 \hline\hline
 UHZ1 & 10.3 & $4 \times 10^7$ & $4\times 10^7$ &  \citet{Bogdan_2023} \\
 \hline
  GHZ9 & 10 & $8 \times 10^7$ & ${\sim} 3\times 10^8$ &  \citet{Kovacs_2024} \\
 \hline
    JADES GN 1146115 & 6.68 & $4 \times 10^{8} $ & $8.3 \times 10^{8}$ & \citet{Juodzbalis_2024} \\
 \hline
 CEERS 1670 & 5.242 & $1.3 \times 10^7$ & $<6\times 10^9$ &  \citet{Kocevski_2023} \\ 
 \hline
 CEERS 3210 & 5.642 & $0.9-4.7 \times 10^7$ & $<6\times 10^{10}$ &   \citet{Kocevski_2023} \\ 
\hline

  CEERS 1019 & 8.679 & $9 \times 10^6$ & $3 \times 10^{9}$ &   \citet{Larson_2023} \\ 
\hline
  GN-z11 & 10.6 & $1.6 \times 10^6$ & $8 \times 10^{8}$ &   \citet{Maiolino_2023} \\ 
\hline

  COSW-106725 & 7.65 & $\geq 6.4 \times 10^8$ & $ 8.3\times 10^{11}$ &    \citet{Lambrides_2023} \\ 
\hline

  Abell2744-QSO1 & 7.0451 & $3 \times 10^7$ & $< 1.4 \times 10^{9}$ &   \citet{Furtak_2023} \\ 
\hline

  PEARLS/NEP-21567 & 14.1& $3.6\times 10^5$ & ?  &    \citet{Nabizadeh_2023} \\ 
\hline
PEARLS/NEP-22802 & 8.2 & $1.5 \times 10^6$ & ? &   \citet{Nabizadeh_2023} \\ 
\hline
    J0100+2802  & 6.327 & $1.15 \times 10^{10}$ & $< 3.38 \times 10^{11}$   & \citet{Yue_2023} \\ 
\hline
    J0148+0600 & 5.977 & $7.79 \times 10^{9}$ & $5.49 \times 10^{10}$  &  \citet{Yue_2023} \\ 
\hline
    J1030+0524   & 6.304 & $1.53 \times 10^{9}$ & $<4.46\times 10^{10}$  &\citet{Yue_2023} \\  
\hline
    J159–02  & 6.381 &   $1.24 \times 10^{9}$ & $1.38 \times 10^{10}$ &   \citet{Yue_2023} \\ 
\hline
    J1120+0641  & 7.085 &   $1.19 \times 10^{9}$ & $6.45 \times 10^{9}$   &\citet{Yue_2023} \\ 
\hline
    J1148+5251  & 6.422 &    $4.36 \times 10^{9}$ & $8.5 \times 10^{10}$ & \citet{Yue_2023} \\ 
\hline
    UNCOVER-20466  & 8.50 &    $1.47 \times 10^{8}$ & $5 \times 10^{8}$ & \citet{Kokorev_2023} \\ 
\hline
    J0371+4459 &  5.01 & $5\times 10^{9}$ & $\leq 5 \times 10^{10}$ & \citet{Stone_2023} \\ 
\hline
    J1340+2813 & 5.36 & $6.3\times 10^{9}$ & $\leq 6.3 \times 10^{10}$ & \citet{Stone_2023} \\  
\hline
    J2239+0207 & 6.25 & $\geq 3.5 \times 10^{8} $ & $\leq 2.5 \times 10^{10}$ & \citet{Stone_2023b, Stone_2023} \\  
\hline
    J2236+0032 & 6.40 & $1.4 \times 10^{9} $ & $1.3 \times 10^{11}$ & \citet{Ding_2023} \\  
\hline
     J2255+0251 & 6.34 & $2.0 \times 10^{8} $ & $3.4 \times 10^{10}$ & \citet{Ding_2023} \\    
     
\hline

\end{tabular}
\caption{A collection of recently discovered high-redshift massive black holes. We share their black hole mass and the stellar mass of their host, if known. UHZ1 has a mass relation of ${\sim 1}$, while several other black holes have $M_{\rm bh}-M_*$ relations ${\sim}0.1$, making these black holes heavy seed candidates.}
\label{table1}
\end{table*}

One of the objects most relevant to this work includes the discovery of a DCBH candidate, detailed in \citet{Bogdan_2023}. Using the Chandra X-ray Observatory, they identify the black hole UHZ1 in a gravitationally-lensed galaxy, behind the cluster lens Abell 2744. Although based only on a few detected X-ray photons, the bolometric luminosity is estimated to be $L {\sim} 5 \times 10^{45}$ erg s$^{-1}$ and assuming Eddington accretion, the implied black hole mass is $4\times 10^7 {\rm M_\odot}$. Comparing this to two different estimates for the surrounding stellar mass, $4\times 10^7 {\rm M_\odot}$ \citep{Castellano_2023} and $7\times 10^7 {\rm M_\odot}$ \citep{Atek_2023}, these observations suggest that if UHZ1 indeed harbors a low-mass SMBH, it is an OMBG with $M_{\rm bh}/M_* {\sim} 1$ \citep{Natarajan_2023, Goulding_2023}, meaning this could be a black hole that originates from direct-collapse, or similar heavy seed models. \citet{Whalen_2023} presents estimates for the radio flux of UHZ1 and estimates the required integration time of 10-100 hr and 1-10 hr for Square Kilometer Array and Very Large Array respectively, which would put even better constraints on this black hole's properties. Given the current measurements, we find that UHZ1 is consistent with the evolution of our DCBHs, shown in Fig.~\ref{fig:coevolution}.

We also highlight other DCBH candidates. The first, detailed in \citet{Kocevski_2023}, is a black hole of mass $1.47\times 10^8 {\rm M}_\odot$. By modeling the spectral energy distribution in optical and near-infrared, they find that the host halo has a stellar mass $<  5\times 10^8 {\rm M}_\odot$. This leads to $M_{\rm bh}/M_* \gtrsim 0.3$. Another over-massive candidate from the JADES survey, detailed in \citet{Juodzbalis_2024}, includes a black hole of mass  $4\times 10^8 {\rm M}_\odot$ which yields a relation of $M_{\rm bh}/M_* \gtrsim 0.4$. Finally, we mention the OMBG candidate GHZ9 at $z{\sim} 10$, with a black hole mass of $8\times 10^7 {\rm M}_\odot$  and a stellar mass of ${\sim}3\times 10^8 {\rm M}_\odot$ \citep{Kovacs_2024,Wang_2024}.

Several additional new SMBHs at redshift $z {\sim} 6$ were identified recently in \citet{Yue_2023}. The six SMBHs discussed in this work have an estimated $M_{\rm bh}/M_*$ ratio similar to $10^{-1}$. While this is almost an order of magnitude larger than the typical SMBH mass relation, given the large masses of these SMBHs, their location in Fig.~\ref{fig:coevolution} suggests that they could still be consistent with light seeds which have experienced rapid growth.  This illustrates the need to find lower-mass SMBHs for the $M_{\rm bh}/M_*$ ratio diagnostic to be useful.

Other recent observations include evidence for black holes that have evolved from light seeds (and may be experiencing super-Eddington accretion) or heavy seeds that have lost their relation. \citet{Kocevski_2023} find two SMBHs, with masses ${\sim} 10^7 {\rm M_\odot}$. They estimate the surrounding stellar mass and find that the $M_{\rm bh}/M_*$ ratio is $10^{-2}$. While this is above location relations ($10^{-3}$), it is no longer possible to determine if this was once an OMBG which has normalised it relation, or if it started as a light seed. \citet{Furtak_2023} find a black hole with a similar relation, while \citet{Lambrides_2023} find a black hole with a lower-limit of $10^{-3}$ on the relation, but potentially much higher. Observations also include a black hole at $z{=} 8.679$, with a mass of ${\sim} 10^7 {\rm M_\odot}$, accreting at $1.2$ times the Eddington limit  \citep{Larson_2023} and a black hole at $z{=} 10.6$, with a mass of ${\sim} 10^6 {\rm M_\odot}$, accreting at ${\sim} 5$ times the Eddington limit \citep{Maiolino_2023}. The estimated stellar mass of these places their $M_{\rm bh}/M_*$ relation at $10^{-3}$, not only well below the OMBG relation, but also below the high-redshift SMBH relation of $10^{-2}$.

While we have focused on the mass relation, DCBHs should also contain unique spectral signatures \citep{Pacucci_2015, Pacucci_2016, Nakajima_2022,Inayoshi_2022}. Using these unique spectral features, \citet{Nabizadeh_2023} finds two DCBH candidates in the PEARLS survey. With future work to determine the stellar mass of their hosts,  their place in the $M_{\rm bh}/M_*$ relation could corroborate their DCBH candidacy. These exciting observations are no doubt just a first glimpse into the future of {\it JWST}'s role in probing the origin of massive black holes at early cosmic times. Our results suggest that we should find many more heavy seeds in the future, which can be safely distinguished from light-seed scenarios.

Recently, \citet{Zhang_2023} have presented and applied their \textsc{Trinity} model to predict halo-galaxy-SMBH connections. They conclude that recent JWST AGNs are broadly consistent with their model. However, they note that UHZ1 is only marginally consistent, and also conclude that it may be in an OMBG phase.

Alternatively, recent work has suggested that these black holes are not inconsistent with local mass relations \citep{Li_2022, Li_2024}. Rather, these black holes may appear over-massive due to a combination of effects including selection biases and measurement uncertainties. Though \citet{Pacucci_2023} argues that recent measurements are significant enough to suggest an intrinsic over-massive relation, future observations and improved measurements will help clarify this possibility.

\section{Conclusions}
\label{sec:conclusion}

The heavy-seed pathway, and specifically the so-called "direct-collapse black hole" scenario producing $10^{5-6}{\rm M_\odot}$ "seed" black holes, remains a promising explanation for the origin of SMBHs of $M \geq 10^9 {\rm M_\odot}$ at redshift $z\sim 6$. At their birth, DCBHs have a uniquely large BH mass to host stellar mass ratio, as emphasised by, e.g. \citet{Agarwal_2013}. S22 measured the lifetime for two DCBH candidates (so-called MMH and LWH, identified by \citealt{Wise_2019}) for which they remain strong outliers in the $M_{\rm bh}/M_*$ relation.  They find that both candidates indeed remain strong outliers down to redshift $z {\sim 8}$ (when they both fall into massive $\sim10^{11}{\rm M_\odot}$ haloes), well into a range where they are potentially detectable by {\it JWST} and sensitive X-ray telescopes. 

In this paper, we followed up on S22 using Monte-Carlo merger trees to analyse the statistics of the over-massive relation lifetime (OMRL) in up to $50,000$ DCBHs across the assembly history of $10^4$ dark matter haloes reaching 
$10^{12}{\rm M_\odot}$ at $z=6$. Using a simple semi-analytic model that accounts for Lyman-Werner irradiation and dynamical heating, we find that each merger tree has $400$--$1200$ DCBH candidates at the time of crossing the atomic-cooling threshold (ACT). We considered two cases, a pessimistic case where only the most irradiated of these candidates from each tree go on to form a DCBH, and an optimistic case where the 5 most irradiated haloes form DCBHs. We find that in both cases, a significant fraction remain strong outliers in the 
$M_{\rm bh}/M_*$ relation, down to redshifts where they become detectable by {\it JWST}. Depending on the minimum mass ratio used to evaluate the OMRL, we find that up to $60\%$ are still outliers at redshift $z=10$, with a comoving number density $\geq 10^{-5} $cMpc$^{-3}$. We expect to find up $0.1-1$ OMBG in each {\it JWST} NIRCam field per unit redshift.

We discussed several recently observed DCBH candidates, compiled in Table\ref{table1}. Most of these objects are still consistent either with a massive seed or a Pop~III stellar-mass seed origin.  However, \citet{Bogdan_2023} has identified a particularly tantalising candidate black hole, UHZ1, at $z=10.3$, for which they inferred $M_{\rm bh}/M_* {\sim}1$. If this object is confirmed to be such a strong outlier, it very strongly favors a massive-seed origin. Future low-mass SMBH discoveries, and their placement in the $M_{\rm bh}/M_*$ relation, will help diagnose the formation pathway of SMBHs with masses $\geq 10^9 {\rm M}_\odot$ at redshift $z\geq 6$.

Finally, as discussed in S22, we note that the $M_{\rm bh}/M_* {\sim}1$ mass-ratio test is not unique to the direct-collapse scenario, but applies to most heavy seeds in general, for which the requirement is to form in a pristine atomic-cooling halo. Our conclusions therefore similarly hold for those scenarios.

\section*{Acknowledgements}
We thank the reviewer and Haojie Hu for their helpful comments and feedback. We thank Robert Feldman and Roberto Maiolino for useful discussions. ZH acknowledges support from NASA grant ATP8ONSSC22K0822 and NSF grants AST-2006176. Merger tree generation and analysis was performed with NSF’s XSEDE allocation AST-120046 and AST-140041 on the Stampede2 resource. The freely available plotting library matplotlib \citep{Hunter_2007} was used to construct the plots in this paper.

\section*{Data Availability}
The code used to analyze the merger trees and generate figures for this manuscript is available at this \href{https://github.com/mscoggs/dcbh_relation_lifetime}{github repository}.  All other data will be shared on reasonable request to the corresponding author.

\bibliographystyle{mnras}
\bibliography{references}
\bsp

\appendix

\section{DCBH mergers and ejection}
\label{sec:appendix}

In the optimistic case, we have assumed 5 of our DCBH candidates go on to form DCBHs, whose host haloes eventually merge into the final DM halo at $z=6$. We have assumed that the DCBHs within these haloes also merge, doing so instantly. We have ignored the possibility of ejection. Here, we briefly discuss a few merger statistics and the possibility of ejection. 

We approximate the escape velocity for the black hole at the virial radius of the halo, which results in a conservative estimate for escape velocity (where leaving the center of the halo would require more energy), and calculate the recoil velocity following \citet{Baker_2008}. The recoil velocity is dependent on several parameters, namely the ratio of the masses, the angles between the black hole spin vectors, and the binary orbital angular momentum vector. Motivated by \citet{Bogdanovic_2007} who argues that external torques during infall help align the black holes, we assume the black holes are completely aligned and the recoil is only dependent on the binary spin magnitudes, the mass ratio, and the fitting parameters of \citet{Baker_2008}. For each recoil, we randomly draw a spin vector for each black hole from a uniform distribution with $0.0 \leq a_{\rm 1,2} \leq 0.9$ and calculate the escape velocity. 

\begin{figure}
  \includegraphics[width=0.85\columnwidth]{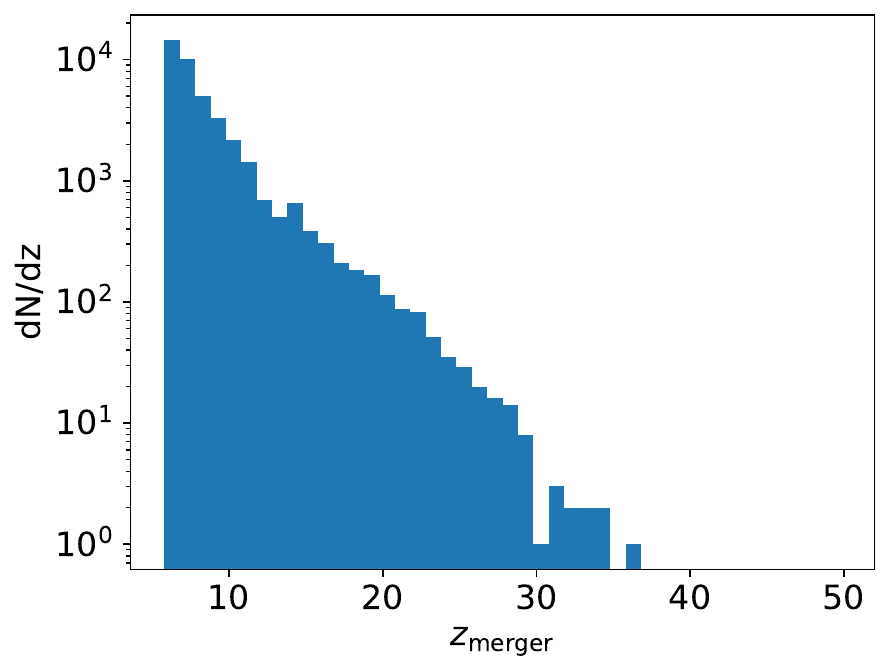}
   \includegraphics[width=0.85\columnwidth]{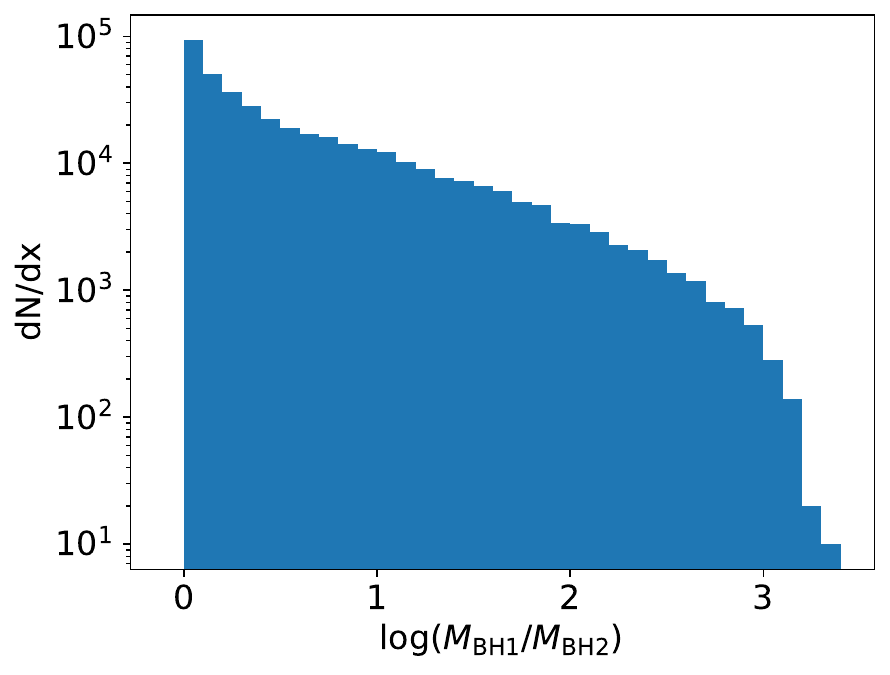}
    \includegraphics[width=0.85\columnwidth]{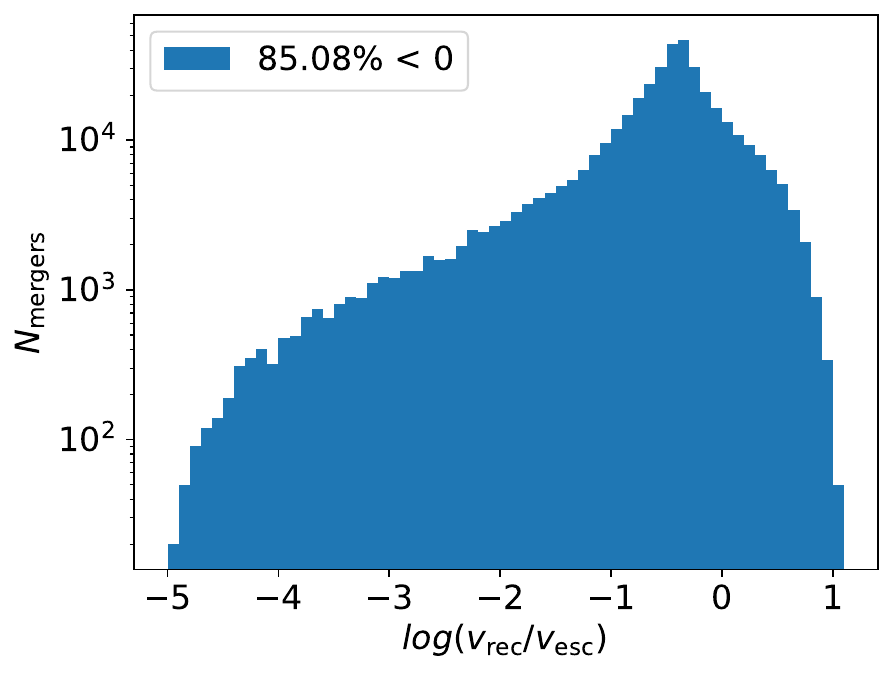}
  \caption{{\it Top:} The redshift distribution of $40,000$ mergers across $10^4$ MC merger trees, hosting a total of 50,000 DCBH candidate haloes (5 per tree). 
  {\it Middle:} The mass ratio of the BH mergers. Black holes are assumed to grow exponentially with an e-folding timescale of $\tau_{\rm cap} = 80$ Myr until they reach a fraction $f_{\rm cap} = 0.1$ of the total baryon mass of the halo, although we find that the mass ratio, and the resulting recoil velocity, is similar for different growth parameters.
  {\it Bottom:} The distribution of the ratio of recoil {\it vs.} escape velocity. We find that $85\%$ of mergers have a recoil velocity less than the escape velocity. Our escape velocity was conservatively estimated by calculating the escape velocity at the virial radius of the halo.}
  \label{fig:merger_figs}
 \end{figure}

We show merger statistics in Fig.~\ref{fig:merger_figs}, calculating the redshift distribution of our mergers (top), the mass ratio of the black holes at those mergers (middle), and a conservative estimate for the ratio of recoil velocity to the escape velocity. As noted in \citet{Volonteri_2006}, \citet{Tanaka_2009}, and \citet{Inayoshi_2020}, these large black holes sitting in the deep potential wells of large dark matter haloes are unlikely to experience recoils because they experience unequal-mass mergers. This leads to a "rich-get-richer" effect where light black holes are likely to be ejected but initially large black holes are typically safely settled into their haloes. Our results agree with this conclusion, where $85\%$ of our mergers have $v_{\rm recoil}/v_{\rm esc} < 1$. A careful account of escape velocity, including dynamical friction and starting with the black hole at the center of the halo, would result in an even larger escape velocity and a higher fraction of mergers where $v_{\rm recoil}/v_{\rm esc} < 1$.

\label{lastpage}
\end{document}